\documentclass[12pt]{article}

\usepackage{epsfig}
\usepackage{amssymb}

\begin{document}

\centerline{\Large {\bf Forecast and Backcast of the Solar Cycles}} 
\vskip 0.5cm
\centerline{\large \bf \em K. M. HIREMATH}
\vskip 0.2cm
\centerline{\large Indian Institute of Astrophysics, Bengaluru-560034, India}
\vskip 0.2cm
\centerline{\large E-mail: hiremath@iiap.res.in}

\begin{abstract}
\noindent Solar cycle is modeled as a forced and damped harmonic oscillator and the
amplitudes, frequencies, phases and decay factors of such a harmonic
oscillator are estimated by non-linear fitting the equation of sinusoidal and
transient parts to the sunspot and irradiance (proxy for the sunspot)
data for the years 1700-2008. We find that:(i) amplitude and frequency
(or period of $\sim$11 yr) of the sinusoidal part remain constant for
all the solar cycles; (ii) the amplitude of the transient part is phase
locked with the phase of the sinusoidal part; (iii) for all the cycles,
the period and decay factor (that is much less than 1) of the transient
part remain approximately constant. The constancy of the amplitudes
and the frequencies of the sinusoidal part and a very small decay
factor from the transient part suggests that the solar activity cycle
mainly consists of a persistent oscillatory part that might be compatible
with long-period ($\sim$22 yr) Alfven oscillations. For all the cycles,
 with the estimated physical parameters (amplitudes, phases and periods)
and, by an autoregressive model, we forecast (especially
for coming solar cycle 25) and backcast (to check whether
Maunder minimum type solar activity exists or not) the solar cycles.
We find that amplitude of coming solar cycle 25 is almost same
as the amplitude of the previous solar cycle 24. We also find that sun might
not have experienced a deep Maunder minimum (MM) type of activity 
during 1645-1700 AD corroborating some of the 
paleoclimatic inferences and, MM type of activity will not
be imminent in near future, until at least 200 years.
\end{abstract}

\section{Introduction}
\vskip 0.5cm
Owing to proximity of the sun, we not only receive the light for sustenance of flora and fauna's life,
but also cyclic and sporadic sun's activities that havoc the electrical grids, satellite
communications and life of the satellites. In addition
to sustenance of life on this Earth, our nearest star
also influences  the climate and environment of the Earth. Recent overwhelming  evidences are
building up that sun indeed influences the earth's climate,
especially Indian Monsoon rainfall (Hiremath and Mandi 2004;
Hiremath 2009; Hiremath, Hegde and Soon 2015 and references there in). Considering these 
important facts, it is essential to understand the origin of solar cycle and
activity phenomena and magnitude of their prediction in future.
Although our understanding of origin of solar cycle and activity phenomena is
far from reality, by learning from the variability
 of sun's long term observed sunspot activity over a century  
scale is very useful and most reliable in projecting the future activity.

Present study aims at this direction and long term variation of sunspot activity
is described as a forced and damped harmonic oscillator. Solution
of such a forced and damped harmonic oscillator is subjected to a non-liner
least square fit  to the long term
sunspot activity data and, amplitudes, frequencies and phases of long period
($\sim$ 22 yr ) oscillations are obtained. With these physical parameters and by the method
autoregression,  sunspot activity after cycle 24 and 
before (cycle 1) the era of regular observations  are obtained.
In the previous study (Hiremath 2006; Hiremath 2008), we have done the similar
analysis. However, present study differs in two aspects: 
(i) back cast of the solar  cycles before the era of so called
1st cycle is obtained and, (ii) recently updated sunspot data as
compiled by ``Royal Observatory of Belgium (http://www.sidc.be/silso/)"
is used for prediction of future cycles 24 and beyond.

Plan of the presentation is as follows. In section 2,
data and method of analyses are presented. Section 3 describes the results
and, brief conclusions are presented in section 4.  

\begin{figure}
\centerline{{\bf Cycle 1} \hskip 21ex  {\bf Cycle 2} \hskip 24ex {\bf Cycle 3}}
\centerline{\psfig{figure=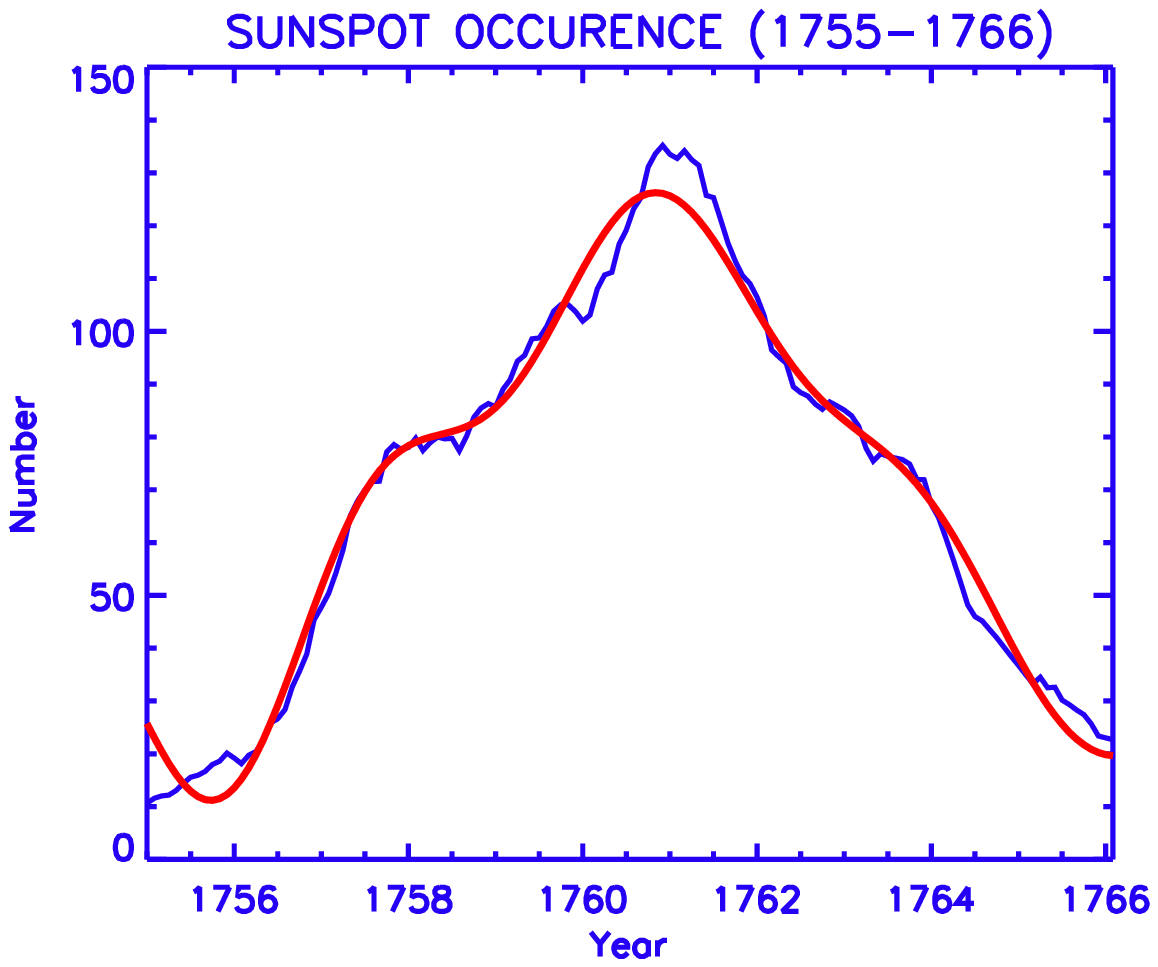, width=6cm,height=6cm}
\psfig{figure=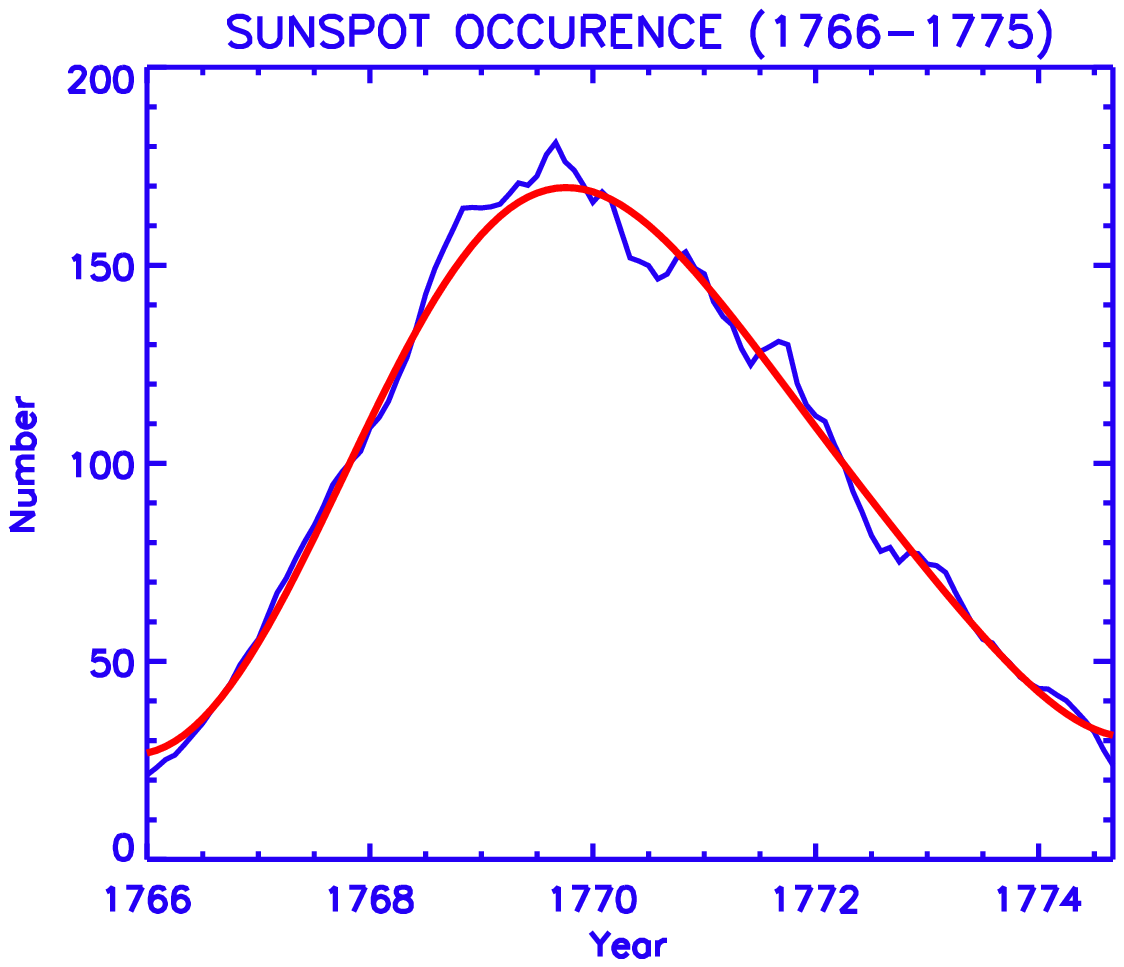, width=6cm,height=6cm}
\psfig{figure=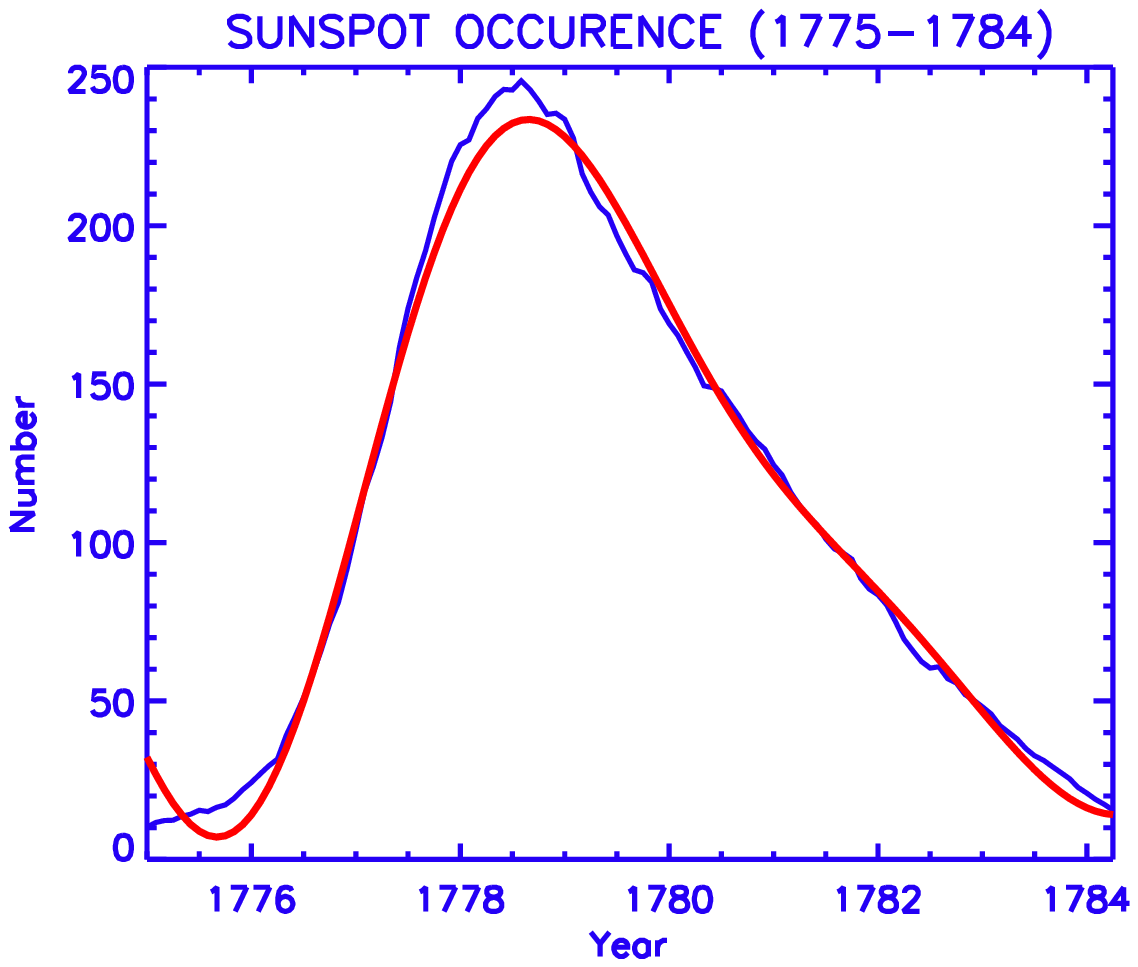, width=6cm,height=6cm}}
\centerline{{\bf Cycle 4} \hskip 21ex  {\bf Cycle 5} \hskip 24ex {\bf Cycle 6}}
\centerline{\psfig{figure=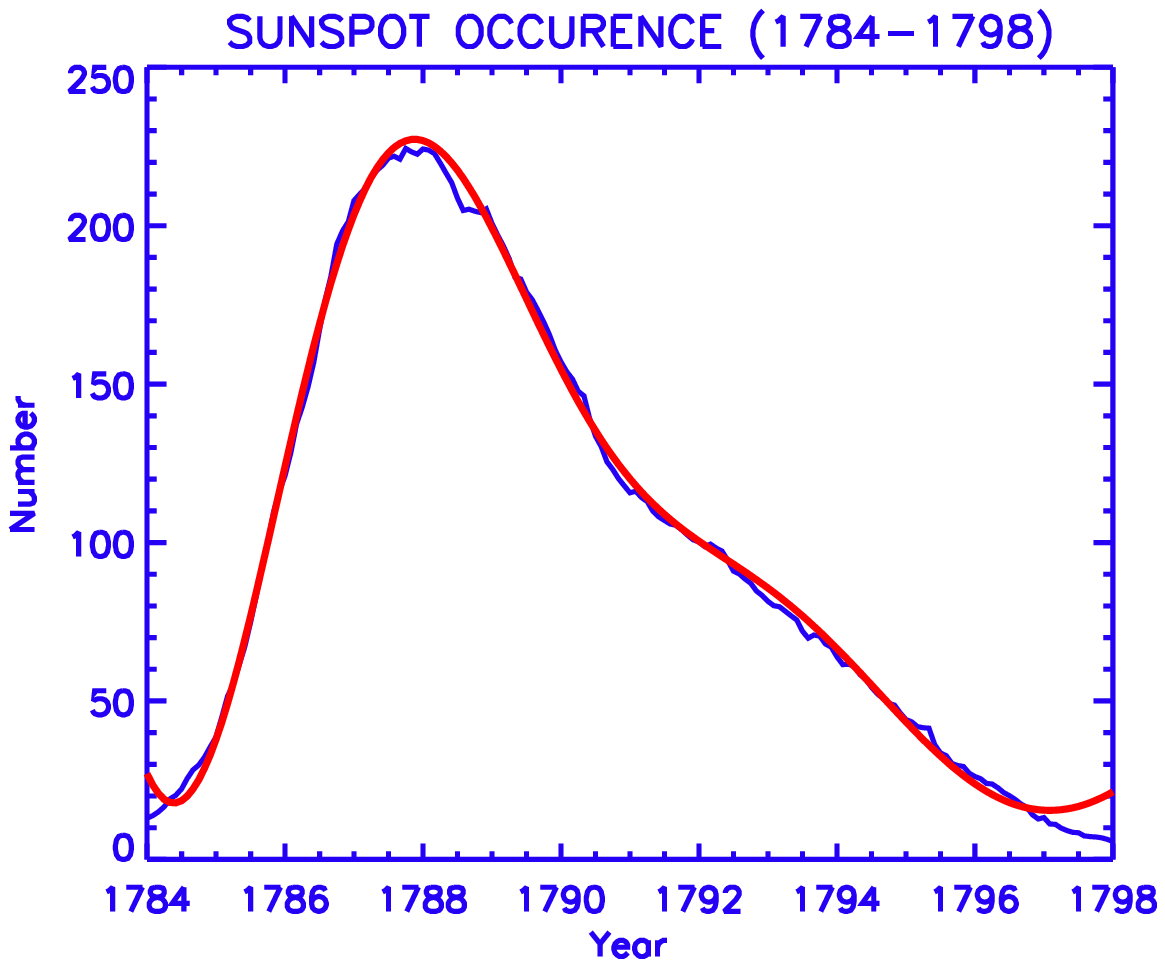, width=6cm,height=6cm}
\psfig{figure=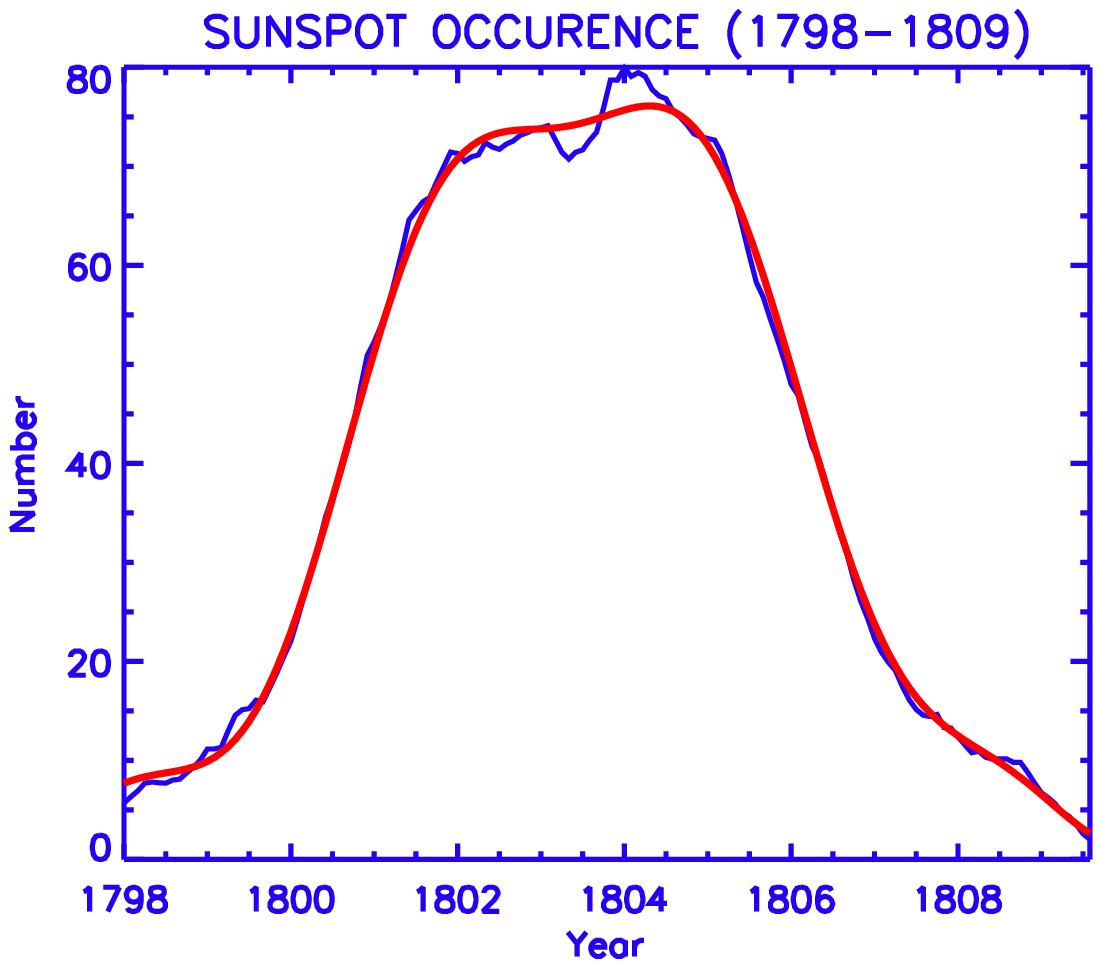, width=6cm,height=6cm}
\psfig{figure=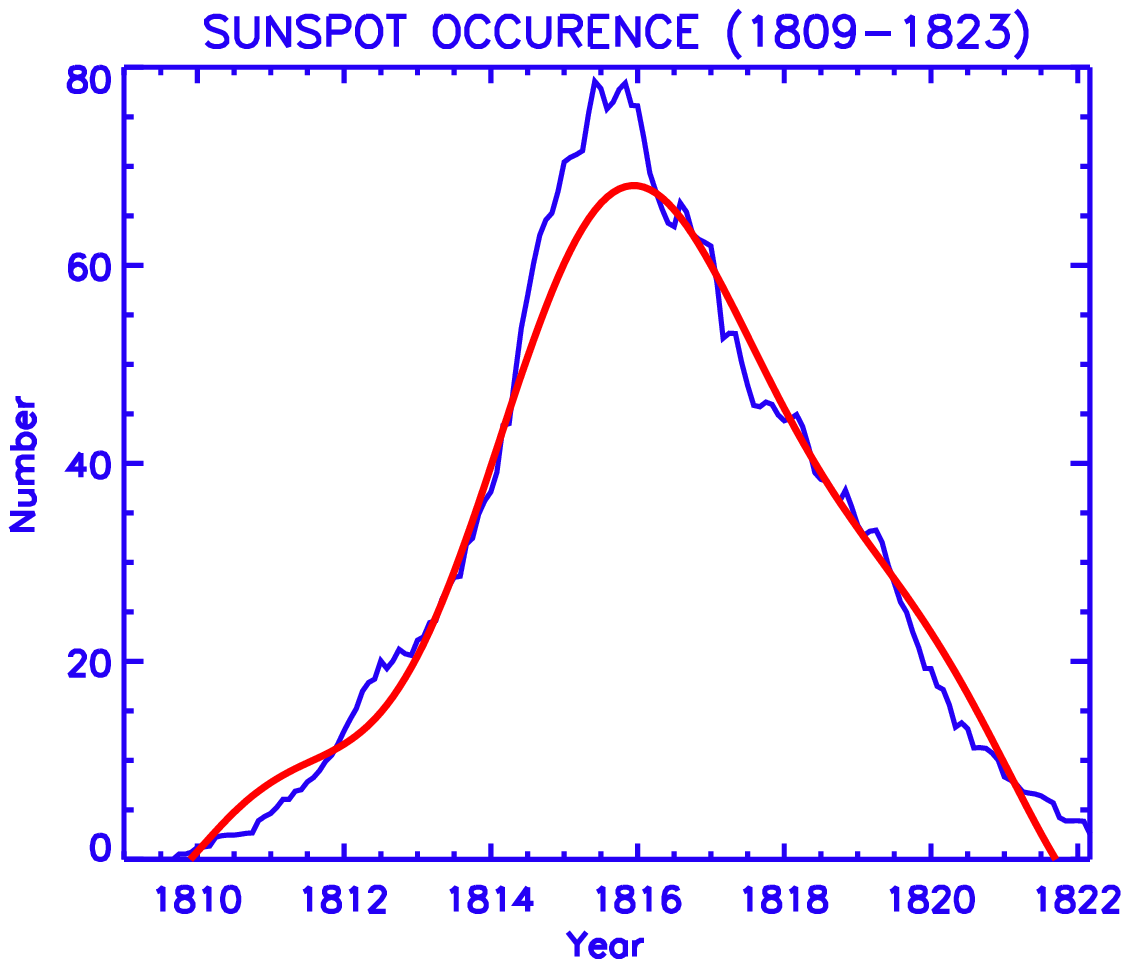, width=6cm,height=6cm}}
\centerline{{\bf Cycle 7} \hskip 21ex  {\bf Cycle 8} \hskip 24ex {\bf Cycle 9}}
\centerline{\psfig{figure=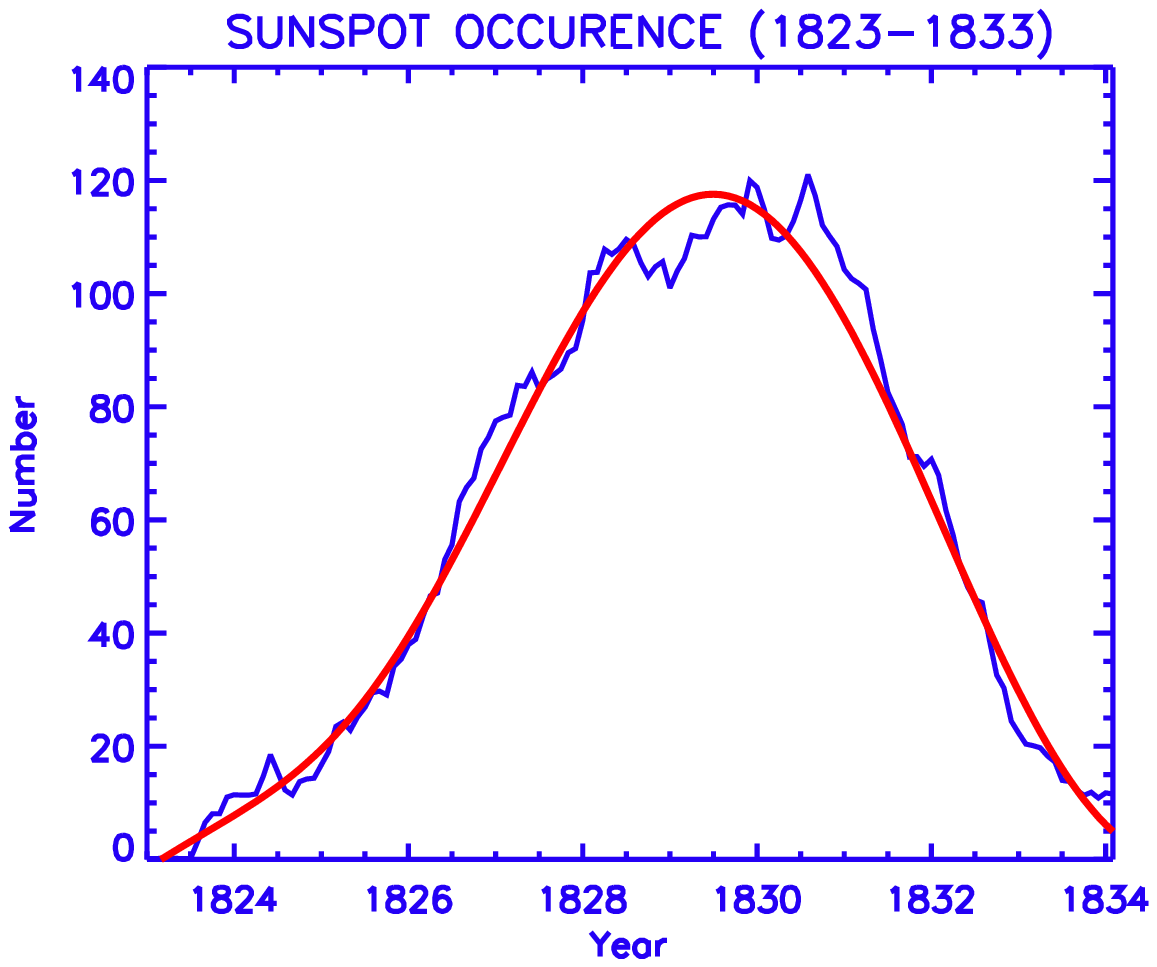, width=6cm,height=6cm}
\psfig{figure=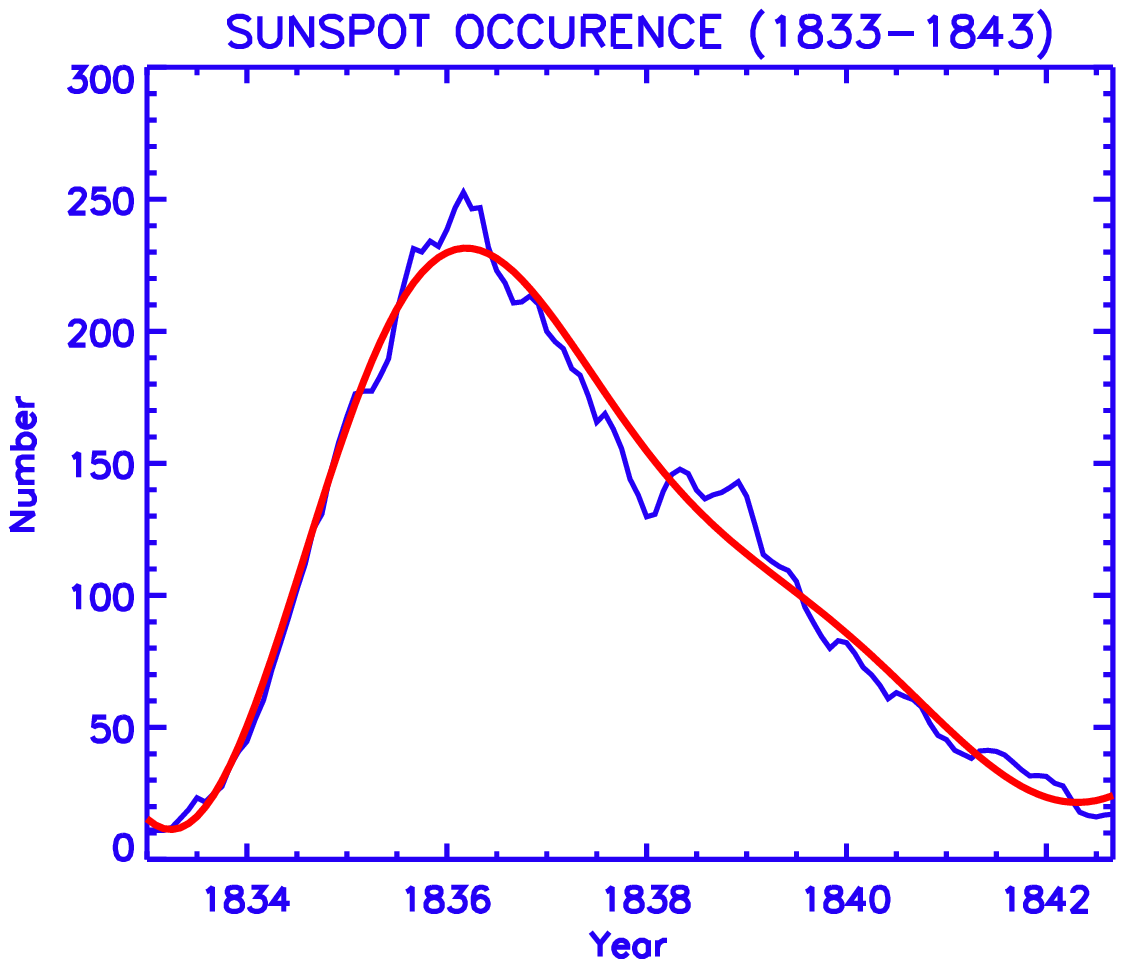, width=6cm,height=6cm}
\psfig{figure=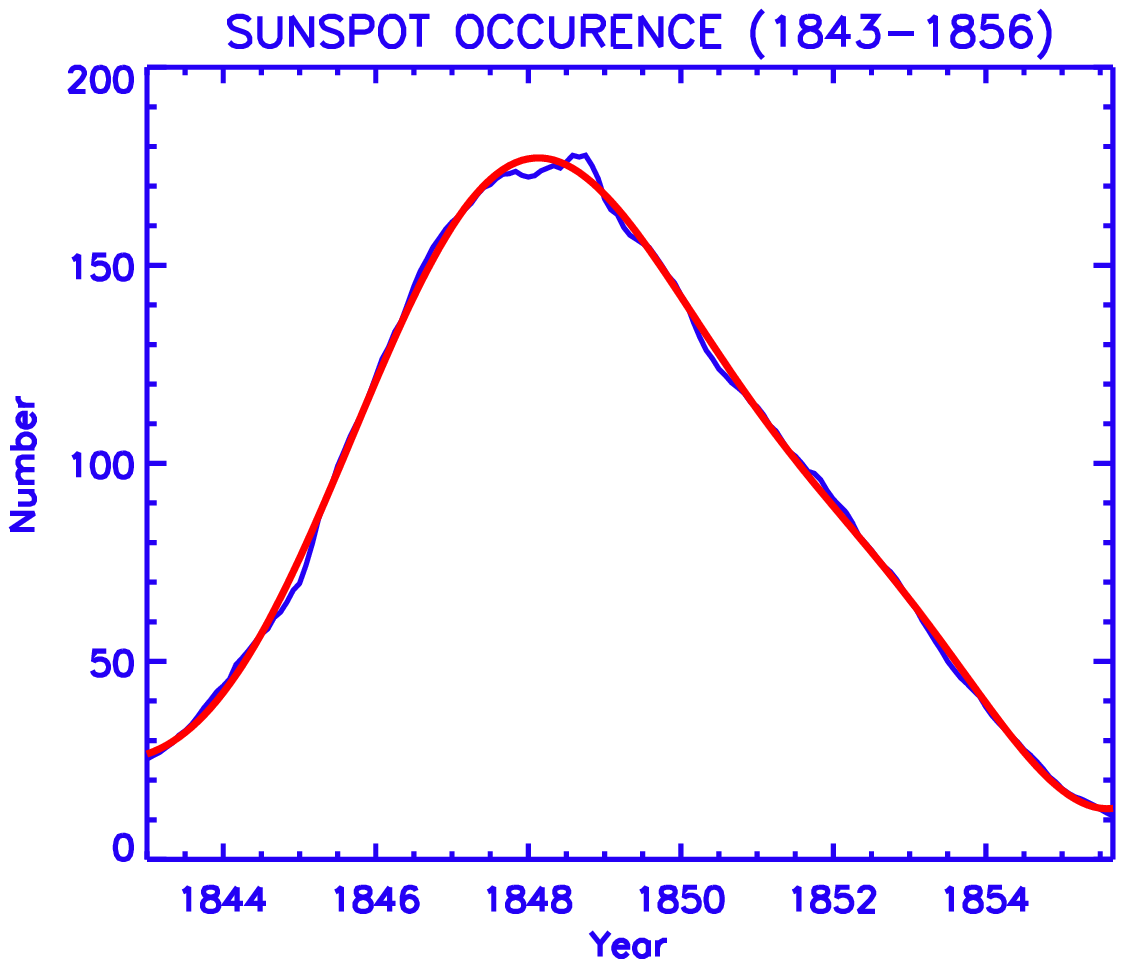, width=6cm,height=6cm}}

\caption{For the solar cycles 1-9, nonlinear least square fit of a solution of forced and
damped harmonic oscillator. Blue continuous line is the observed sunspot data and
red continuous line is obtained from the fit.}
\end{figure}

 \begin{figure}
\centerline{{\bf Cycle 10} \hskip 21ex  {\bf Cycle 11} \hskip 24ex {\bf Cycle 12}}
\centerline{\psfig{figure=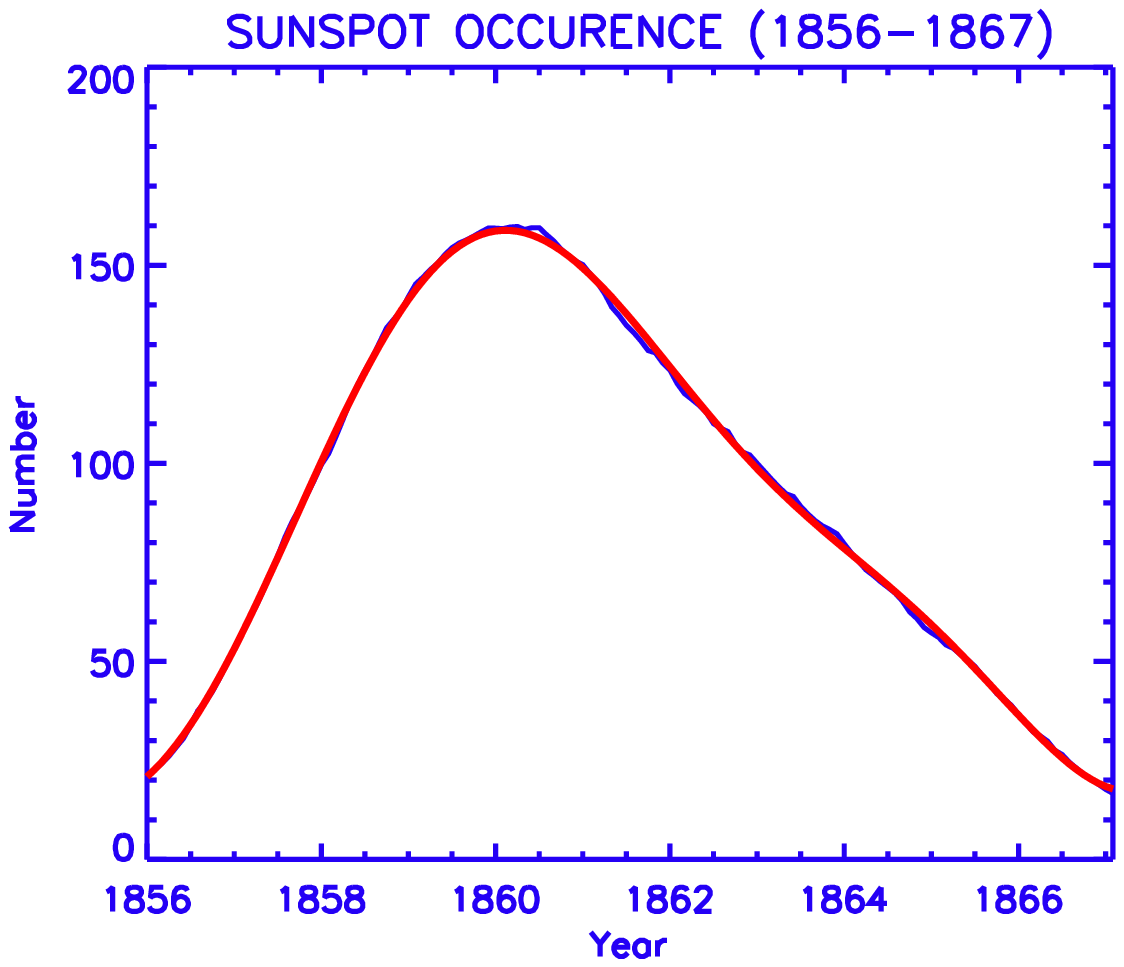, width=6cm,height=6cm}
 \psfig{figure=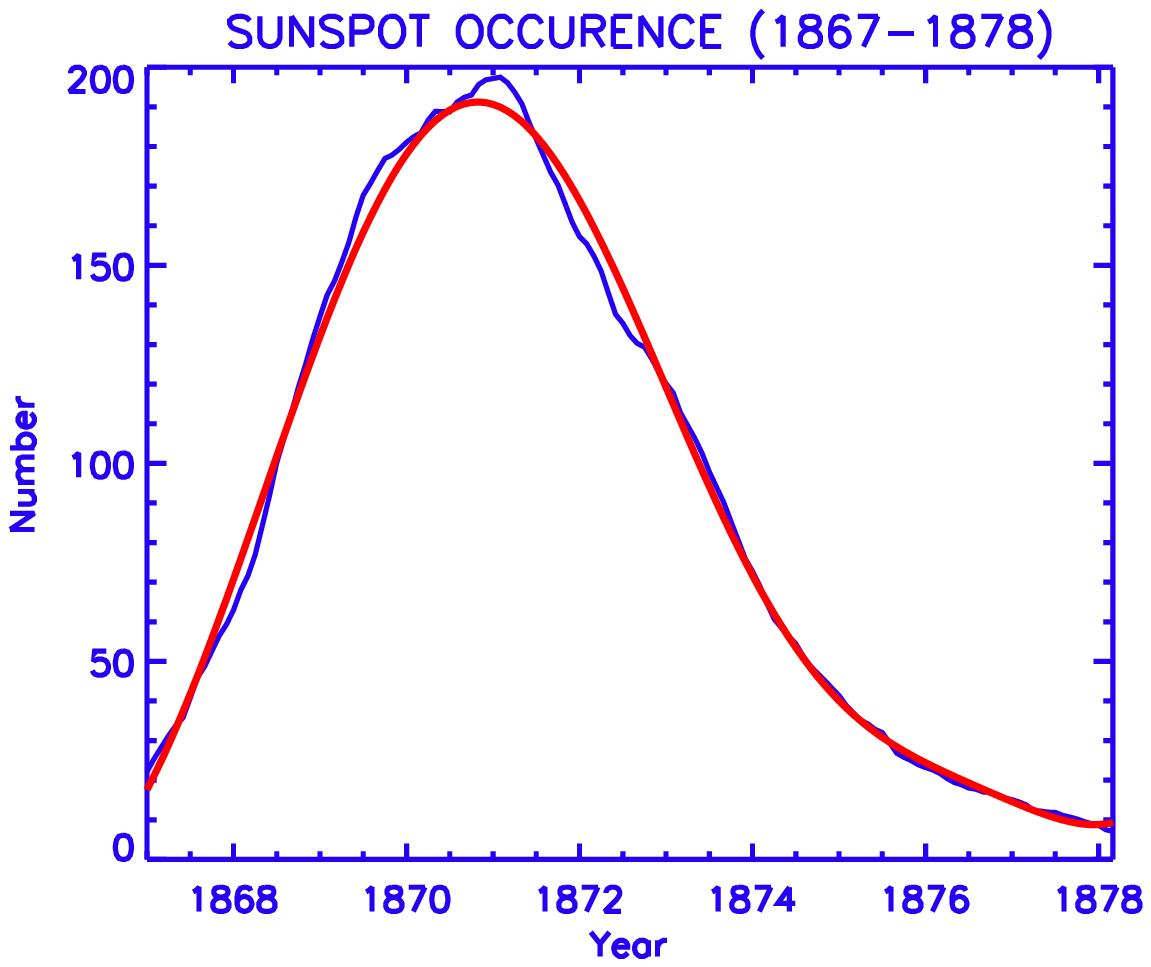, width=6cm,height=6cm}
\psfig{figure=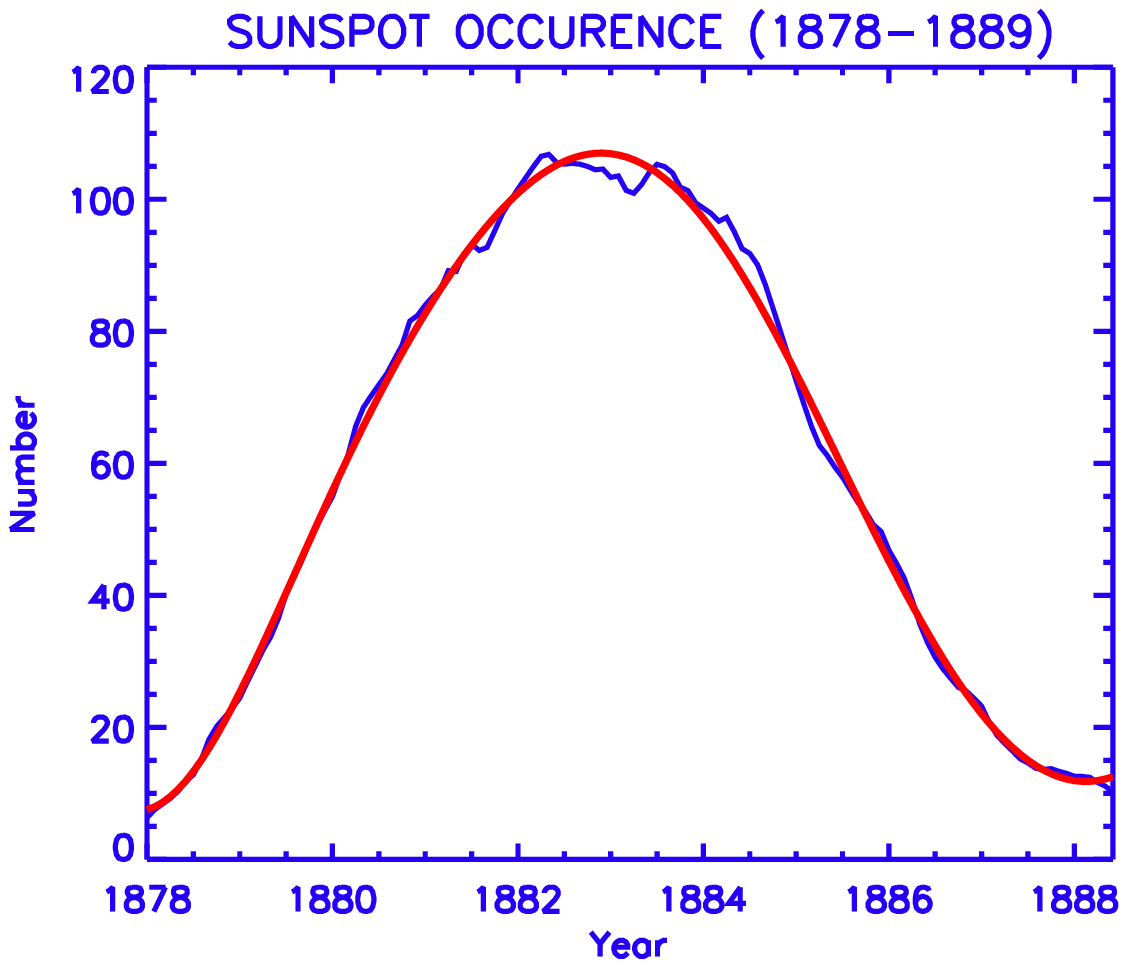, width=6cm,height=6cm}}
\centerline{{\bf Cycle 13} \hskip 21ex  {\bf Cycle 14} \hskip 24ex {\bf Cycle 15}}
\centerline{\psfig{figure=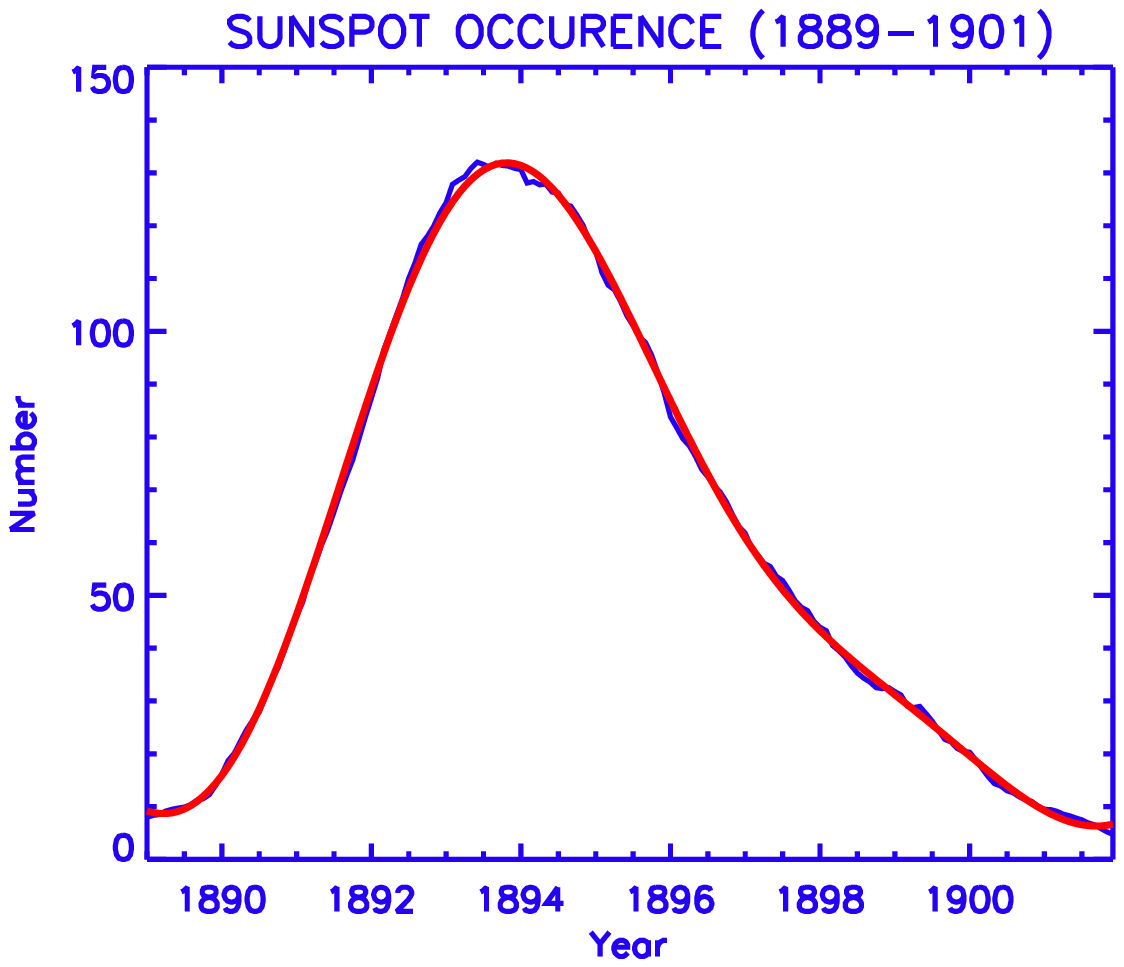, width=6cm,height=6cm}
\psfig{figure=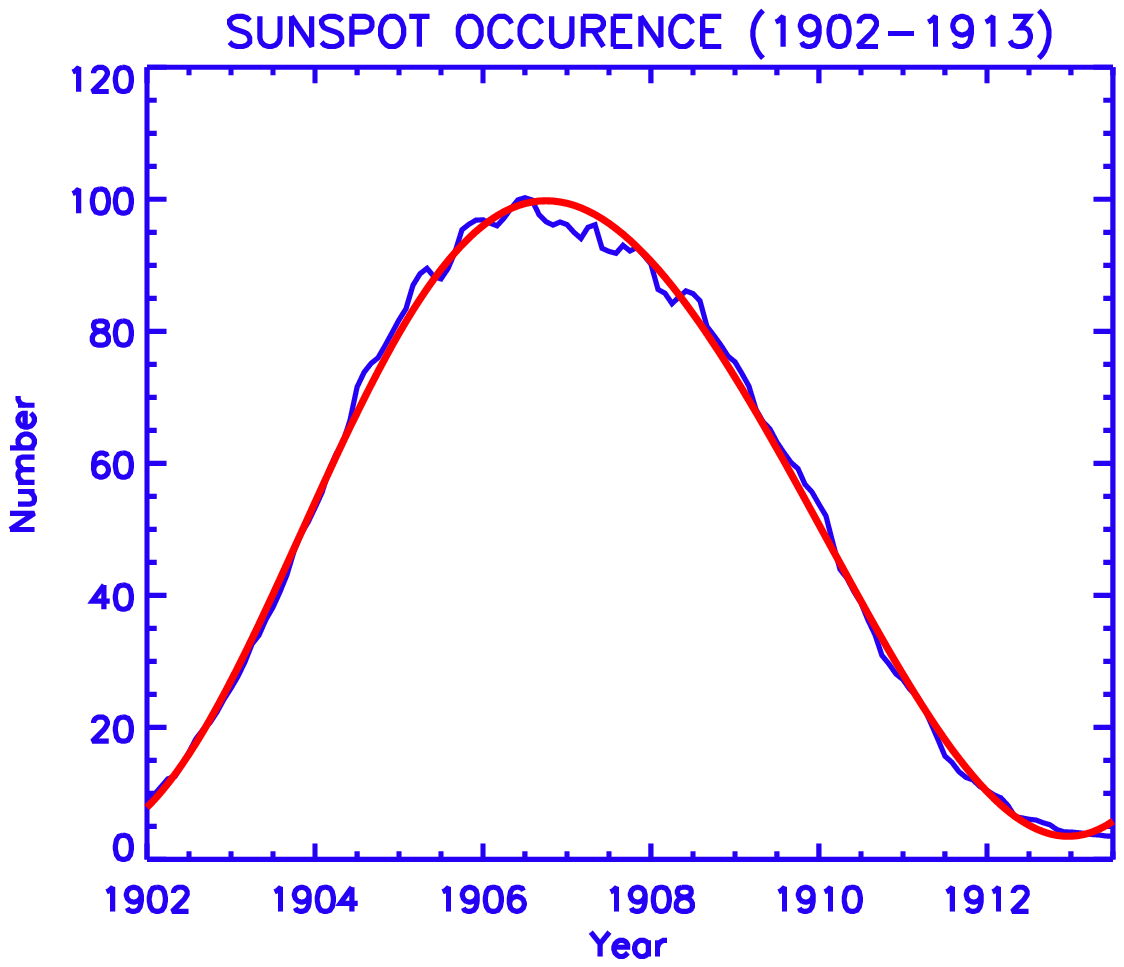, width=6cm,height=6cm}
\psfig{figure=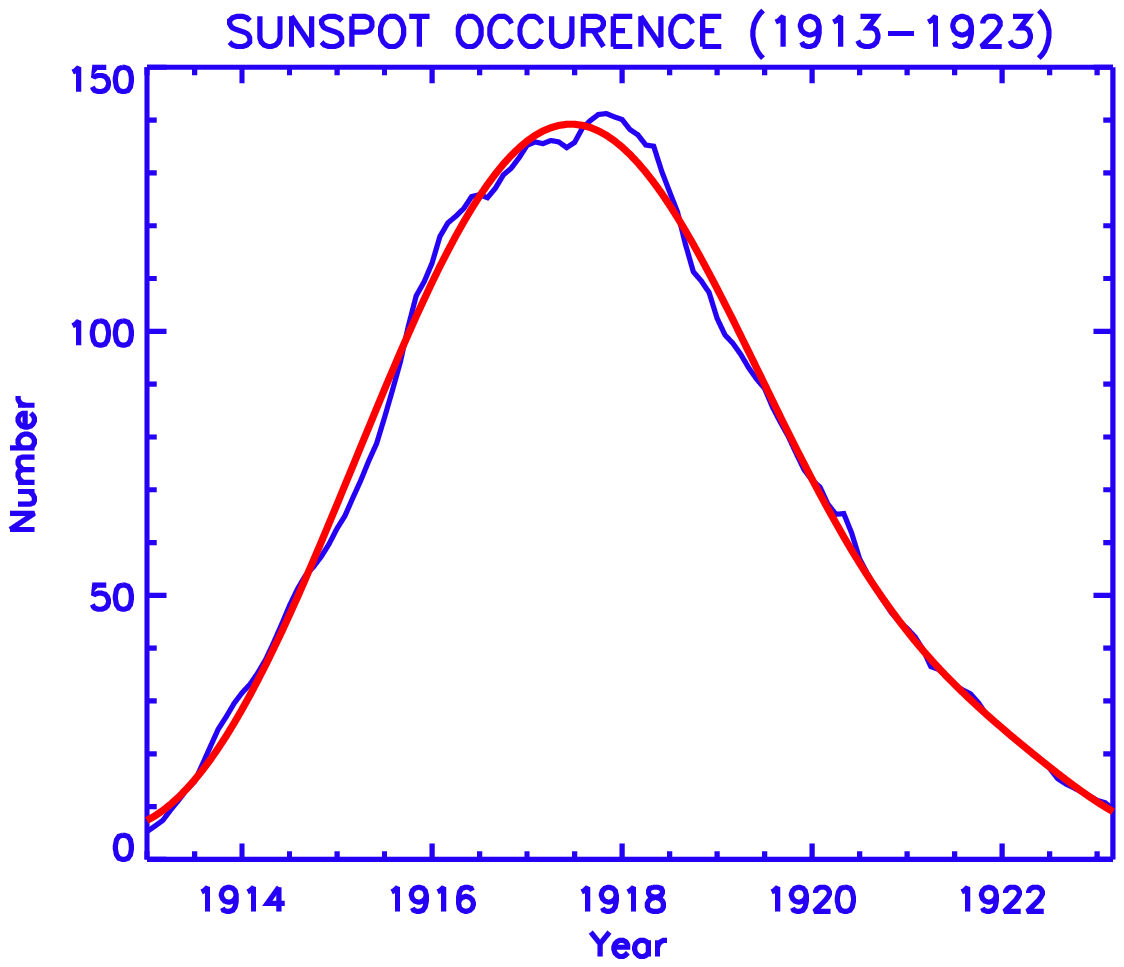, width=6cm,height=6cm}}
\centerline{{\bf Cycle 16} \hskip 21ex  {\bf Cycle 17} \hskip 24ex {\bf Cycle 18}}
\centerline{\psfig{figure=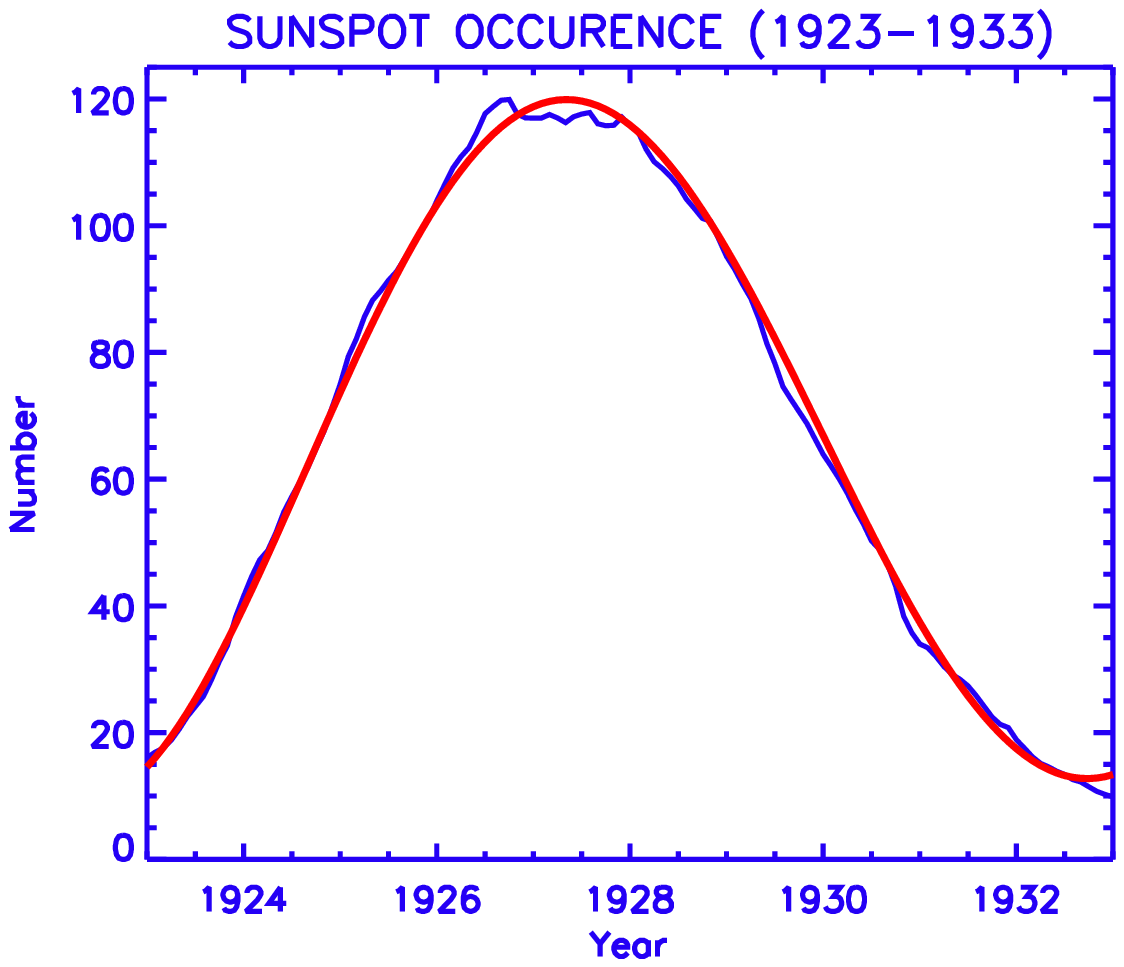, width=6cm,height=6cm}
\psfig{figure=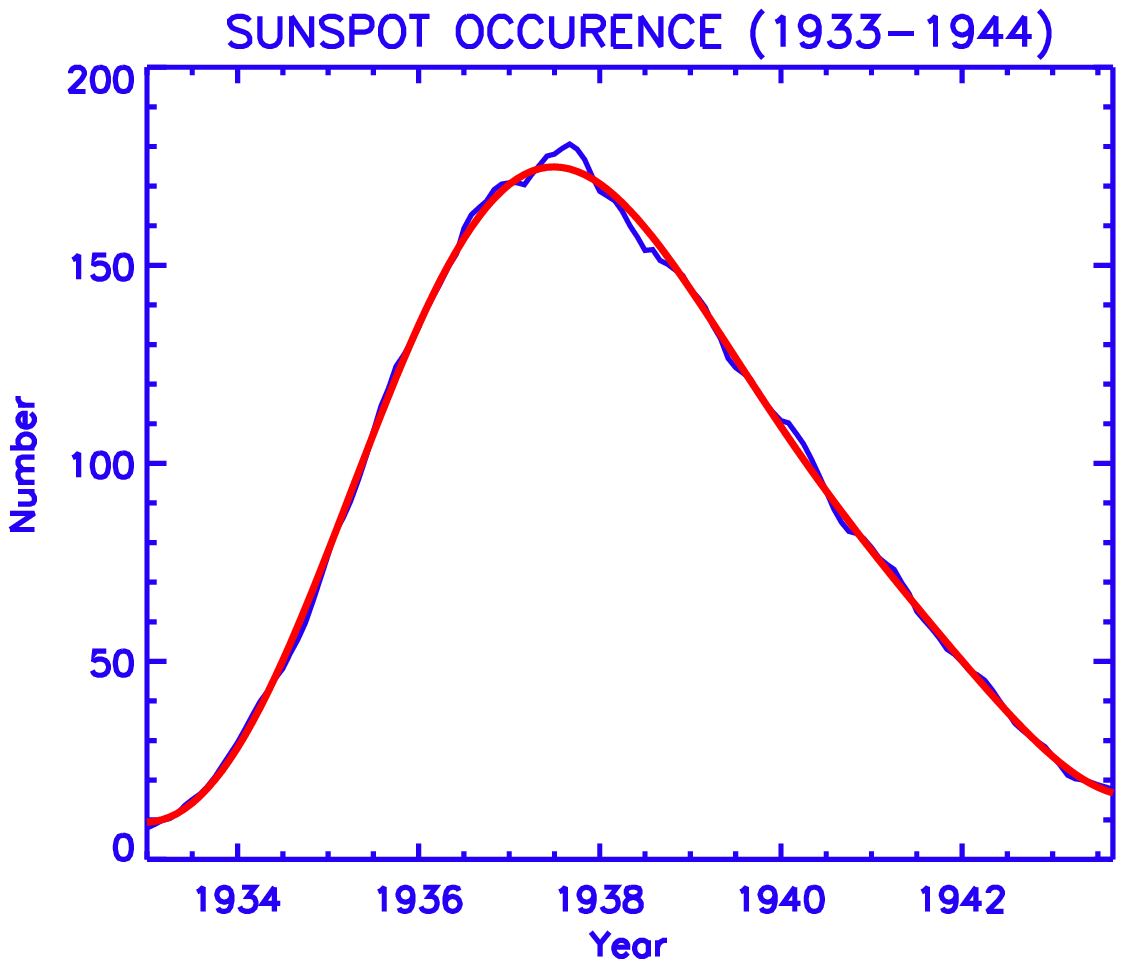, width=6cm,height=6cm}
\psfig{figure=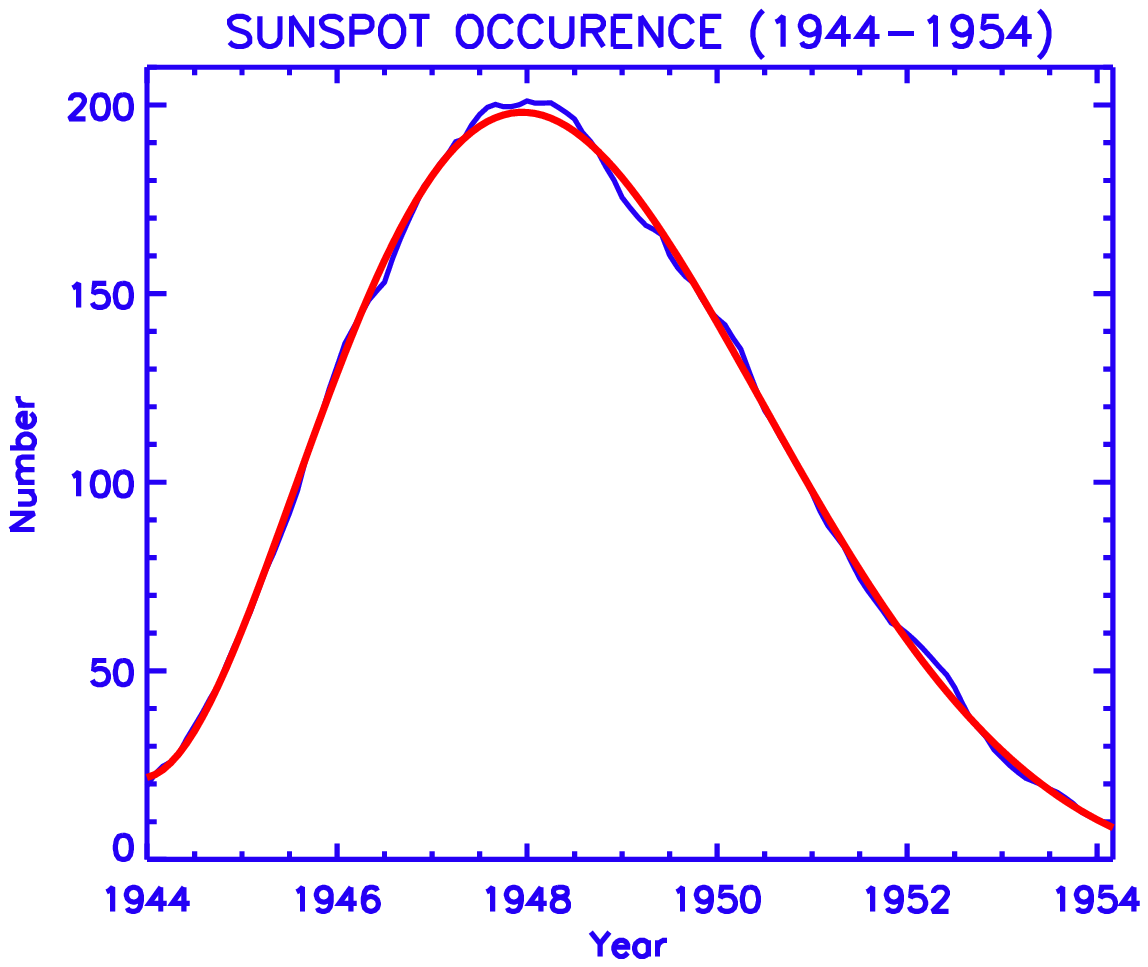, width=6cm,height=6cm}}

\caption{For the solar cycles 10-18, nonlinear least square fit of a solution of forced and
damped harmonic oscillator. Blue continuous line is the observed sunspot data and
red continuous line is obtained from the fit.}
\end{figure}

 \begin{figure}
\centerline{{\bf Cycle 19} \hskip 21ex  {\bf Cycle 20} \hskip 24ex {\bf Cycle 21}}
\centerline{\psfig{figure=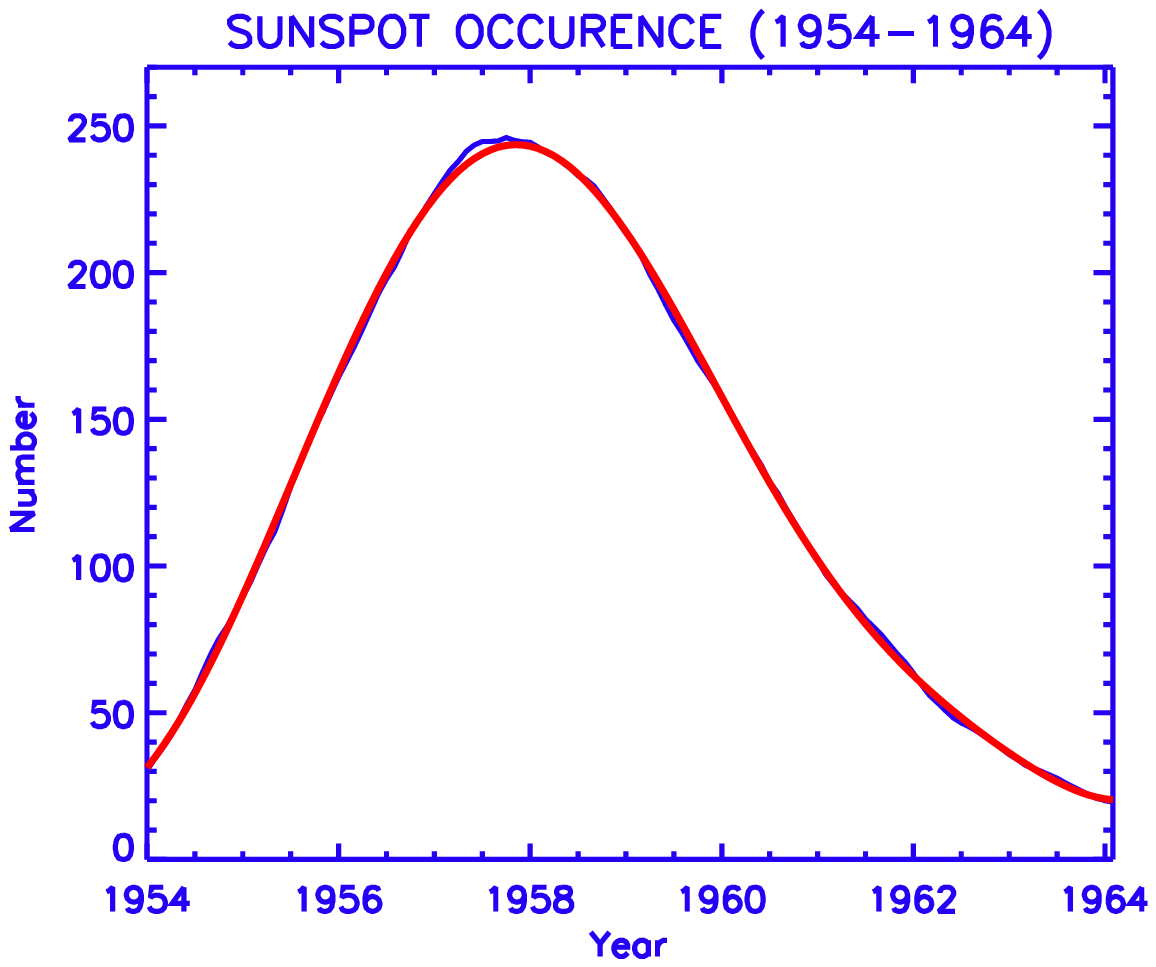, width=6cm,height=6cm}
 \psfig{figure=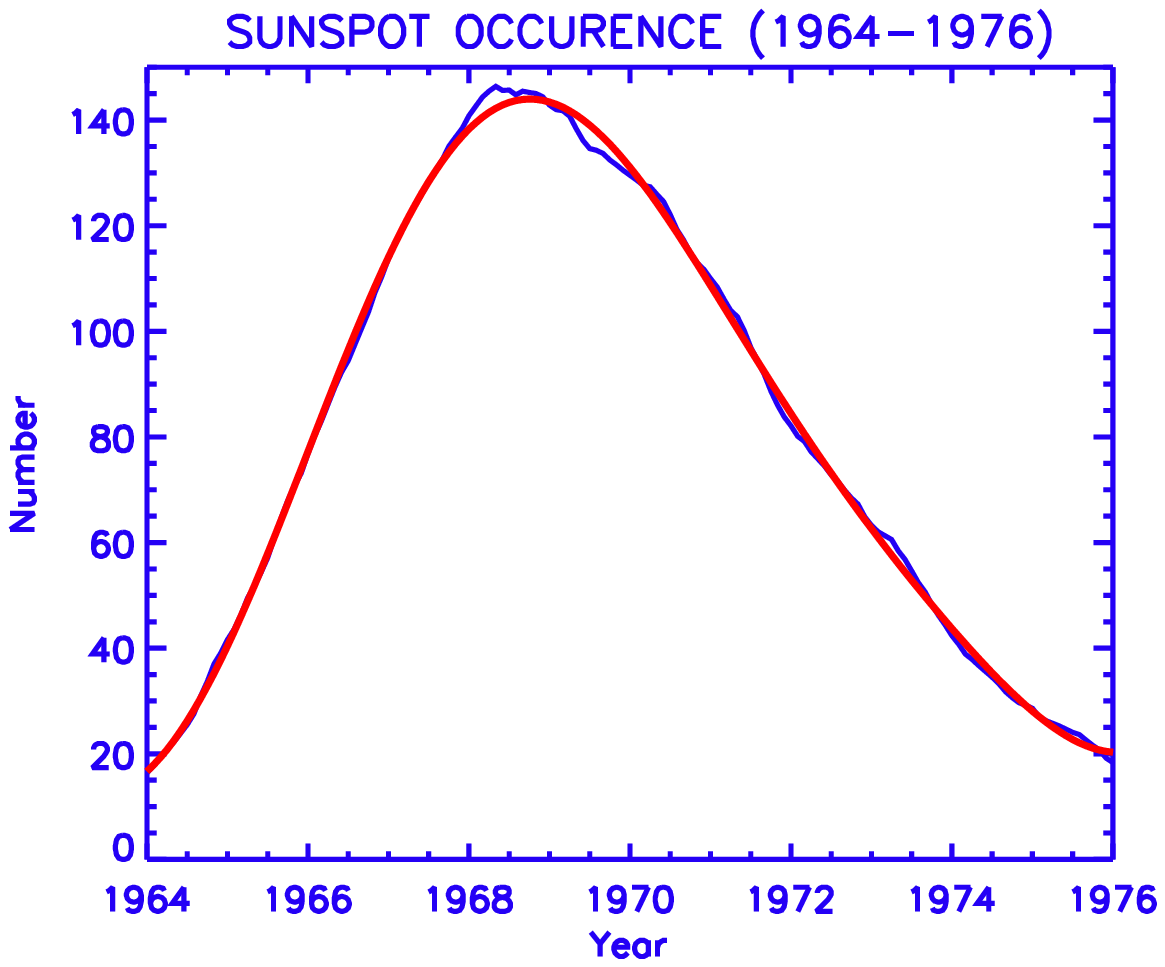, width=6cm,height=6cm}
\psfig{figure=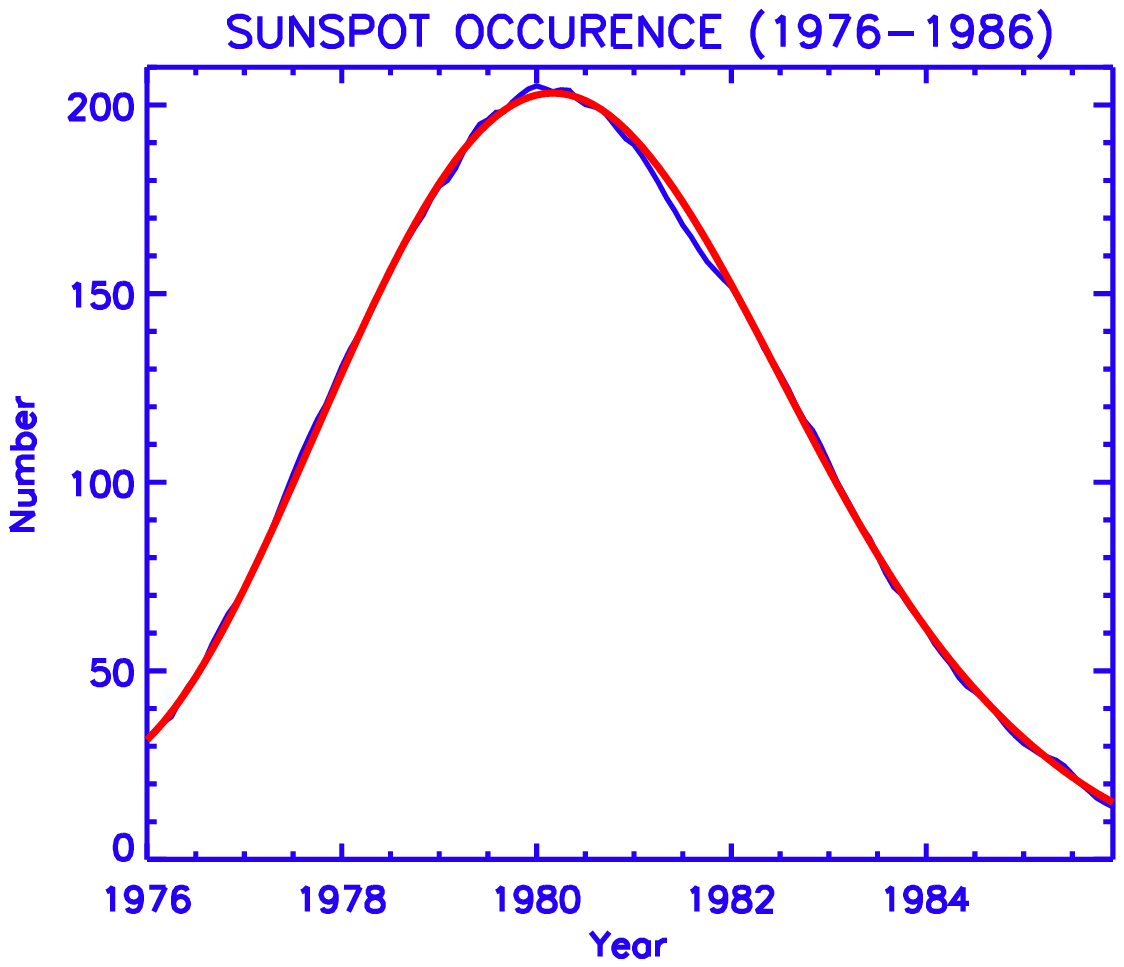, width=6cm,height=6cm}}
\centerline{{\bf Cycle 22} \hskip 21ex  {\bf Cycle 23}}
\centerline{\psfig{figure=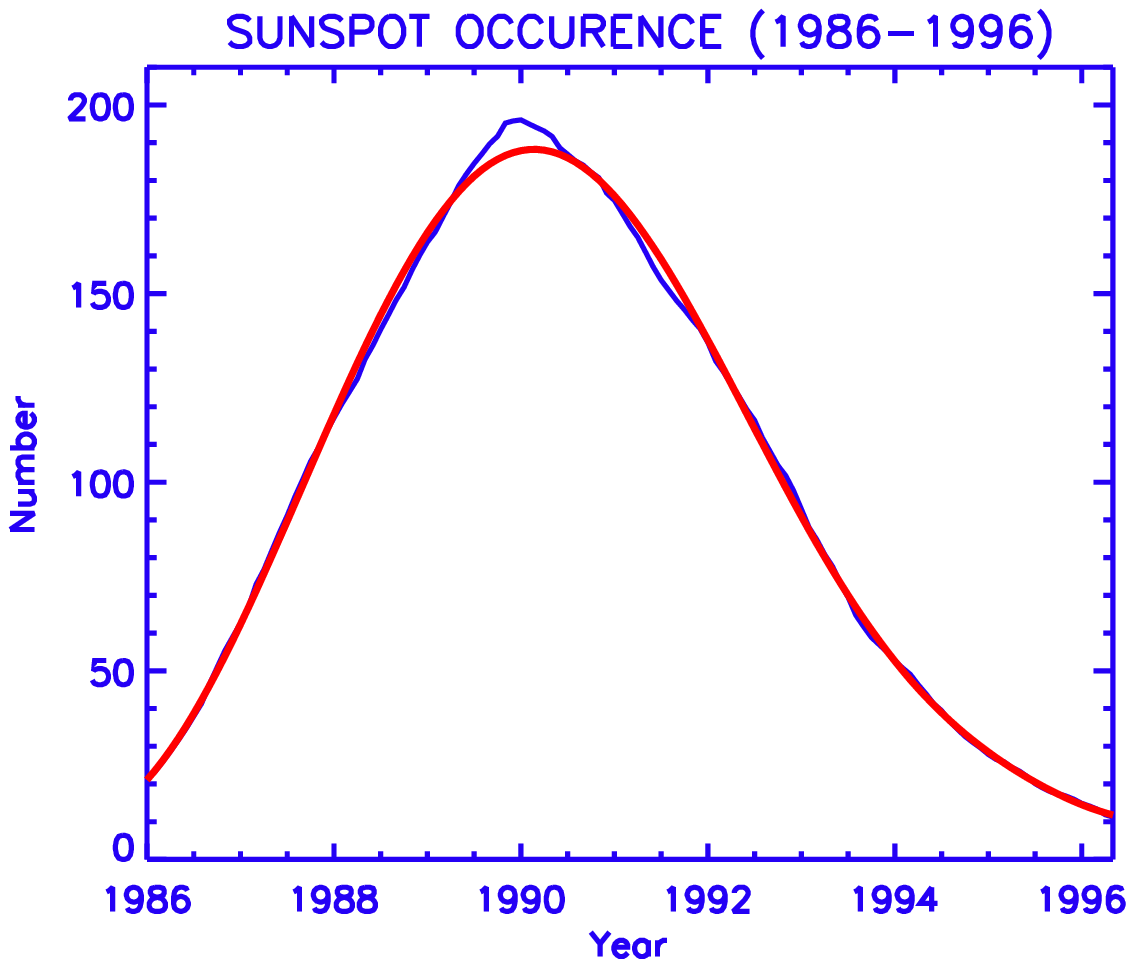, width=6cm,height=6cm}
\psfig{figure=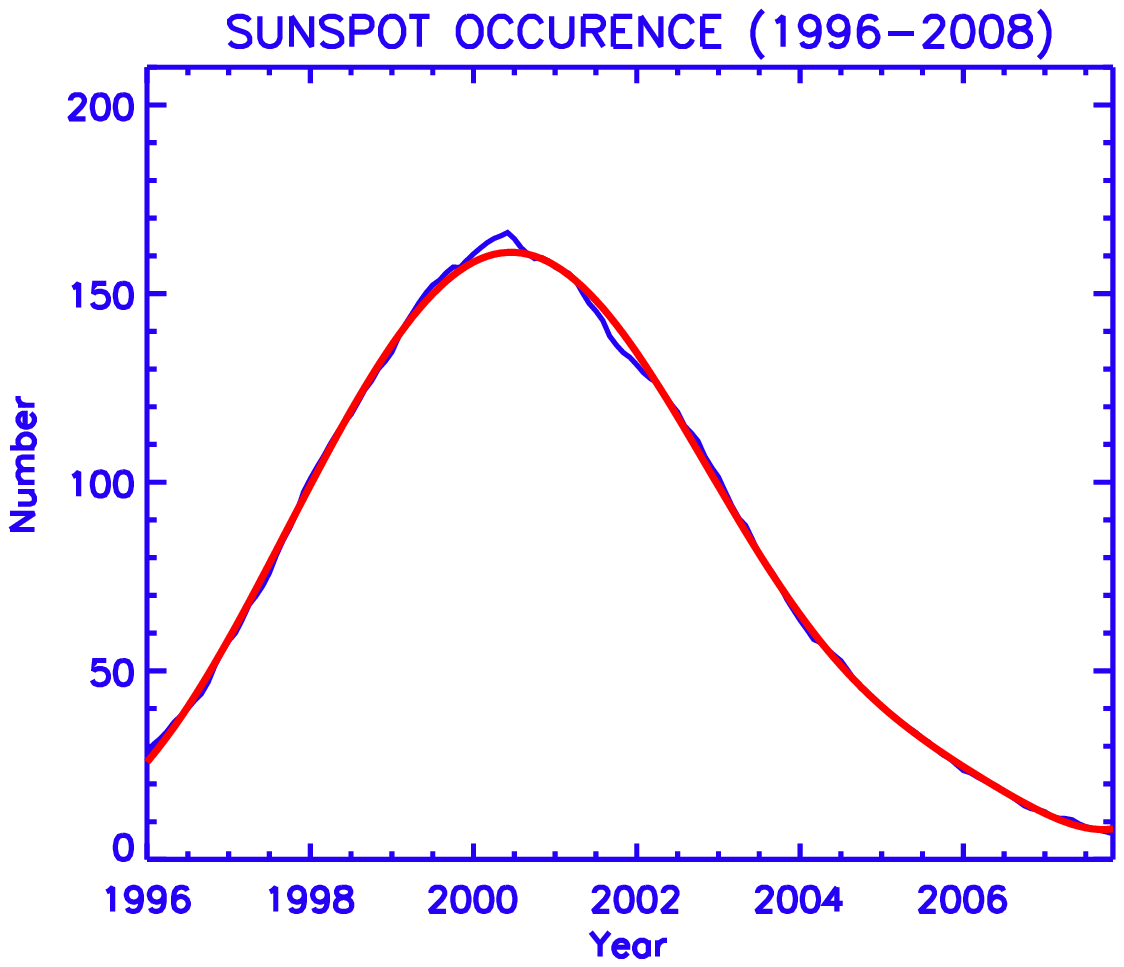, width=6cm,height=6cm}}
\caption{For the solar cycles 19-23, nonlinear least square fit of a solution of forced and
damped harmonic oscillator. Blue continuous line is the observed sunspot data and
red continuous line is obtained from the fit.}
\end{figure}

 \begin{figure}
{\bf Fig-4(a)} \hskip 30ex  {\bf Fig-4(b)}
\centerline{\psfig{figure=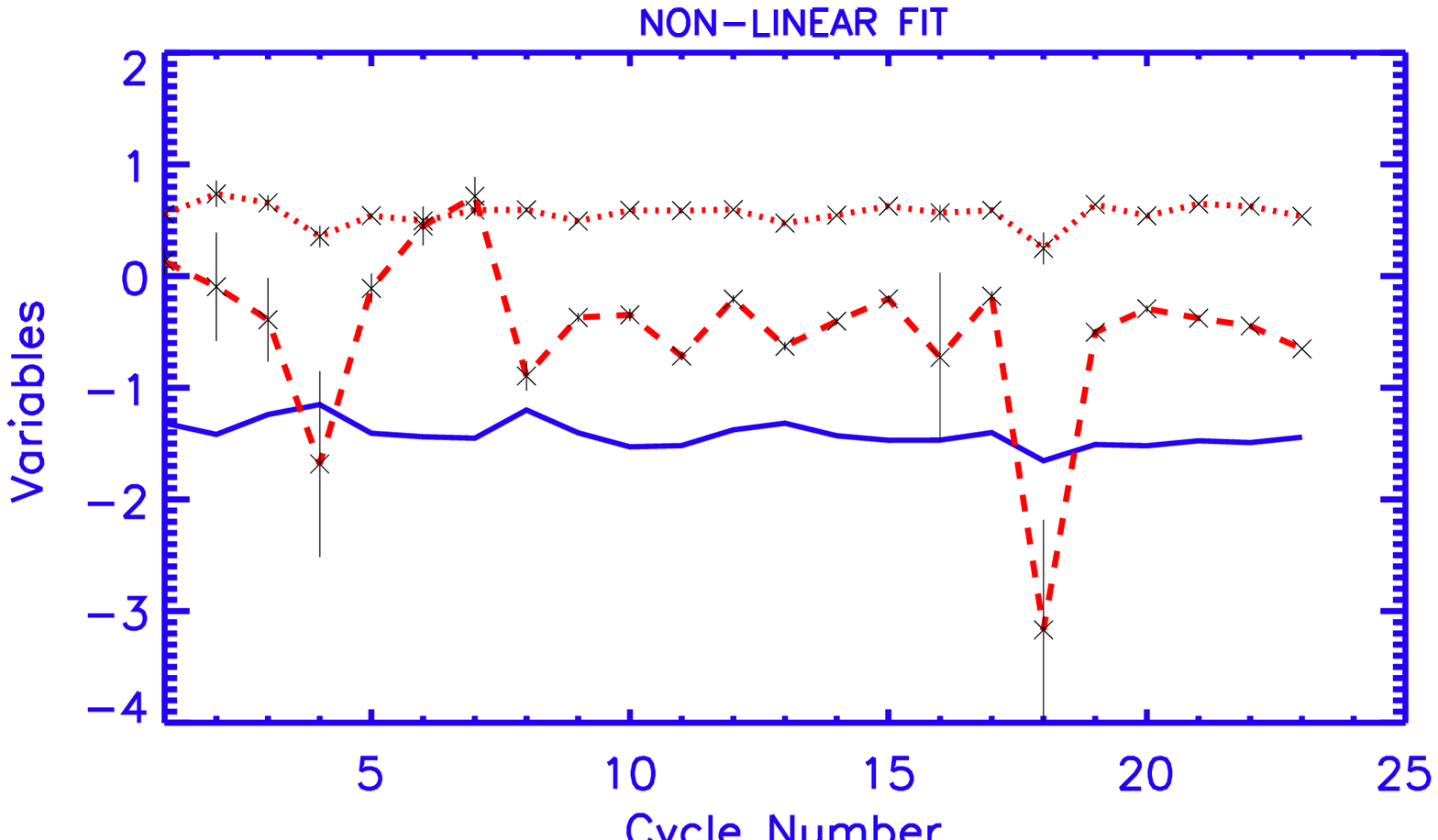,width=8cm,height=6cm}
\psfig{figure=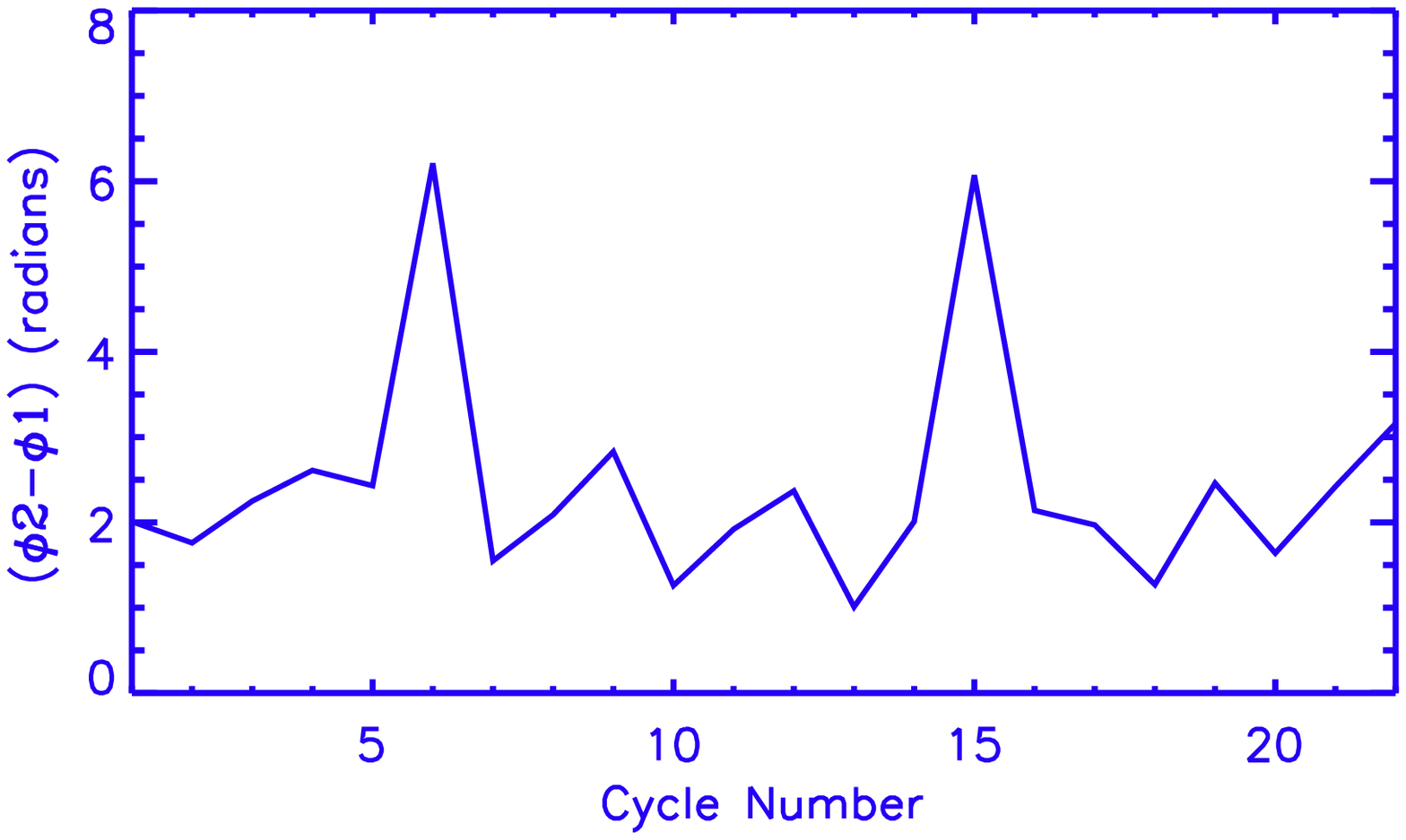, width=8cm,height=6cm}}
{\bf Fig-4(c)} \hskip 30ex  {\bf Fig-4(d)}
\centerline{\psfig{figure=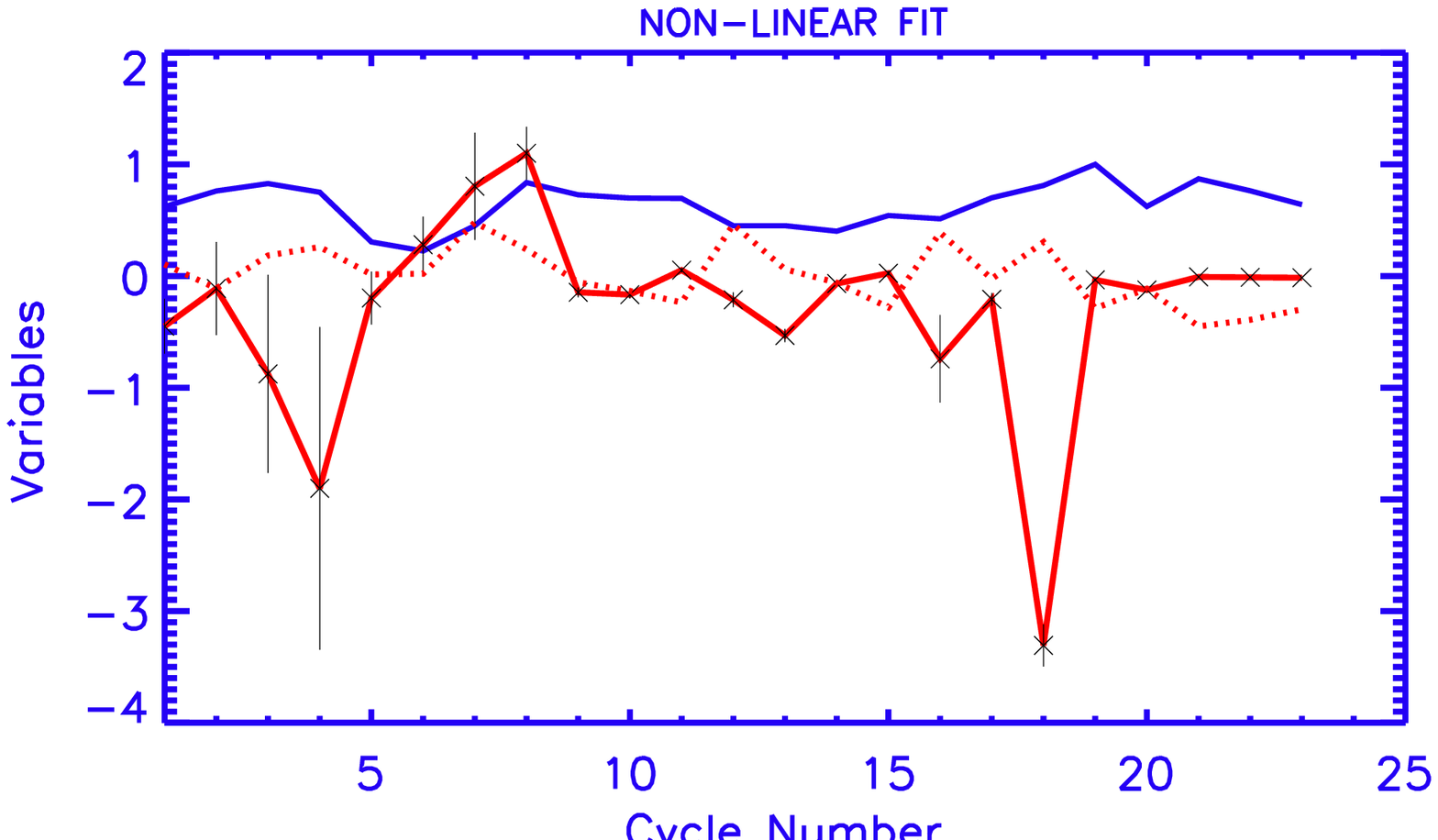, width=8cm,height=6cm}
\psfig{figure=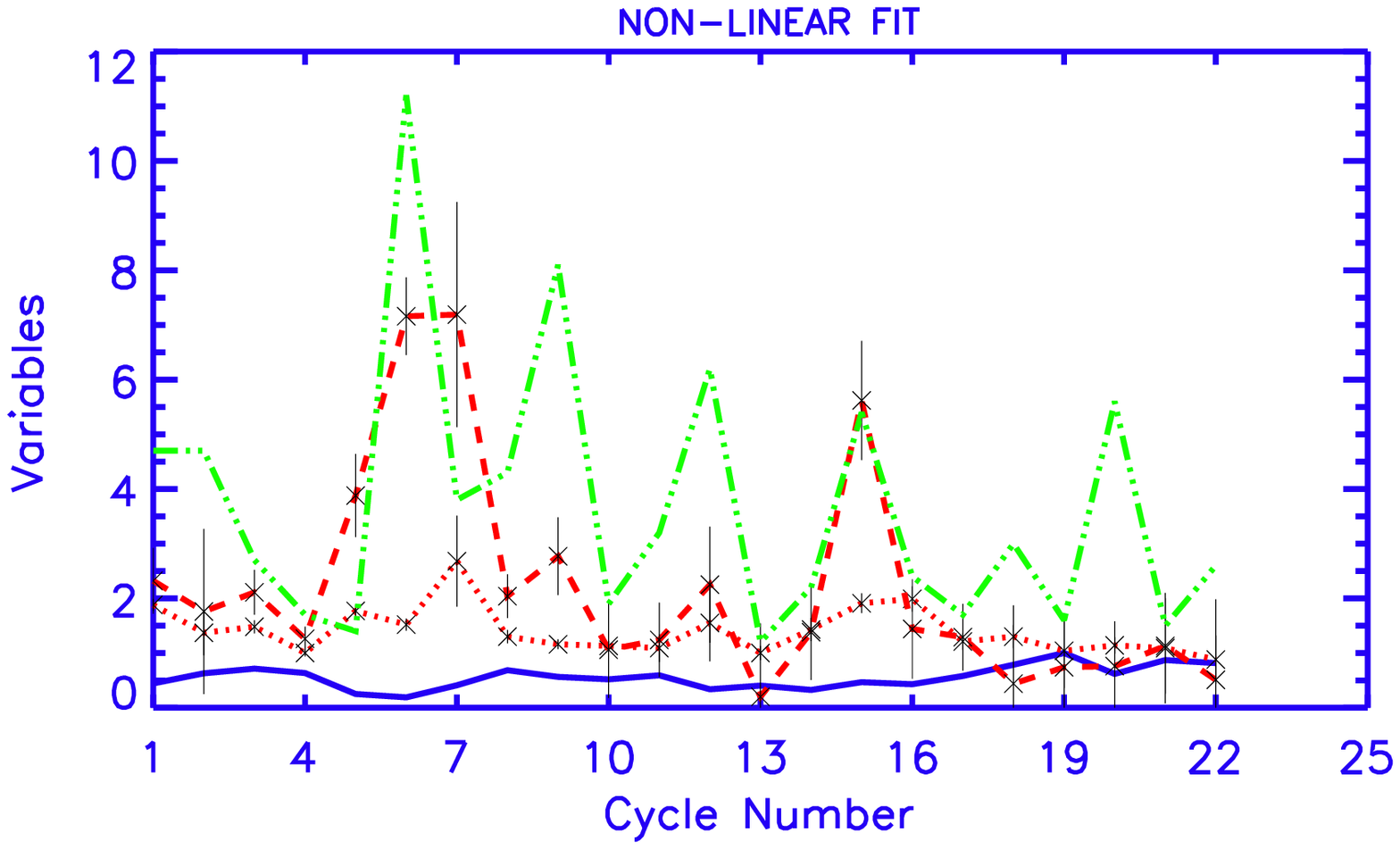, width=8cm,height=6cm}}
\caption{For the solar cycles 1-23, Fig 4(a) illustrates the variation 
of different coefficients (blue continuous line is the amplitude $A_{1}$, 
the dotted line is the frequency $\omega$ and the dashed line is the 
phase $\phi _{1}$) for the steady part. Whereas variation of different physical
parameters of transient part are illustrated in Fig 4(c) and Fig 4(d) respectively.
In Fig 4(c), red continuous line is the amplitude $A_{2}$ and the 
dotted line represents the decay factor $\gamma $ respectively. 
In Fig 4(d), the dotted line is the frequency $\omega '$ and the dashed 
line is the phase $\phi _{2}$. The dash with three dotted line represents 
the values of $\chi ^2$ (a measure of goodness of fit) for each cycle.
Fig 4(b) illustrates the phase difference between steady and transient 
parts. In both the illustrations of Fig 4(c) and Fig 4(d) 
blue continuous line is the cycle mean (normalized to the maximum number in all the solar cycles). }
\end{figure}

\section{Data and Analysis}

 For the years 1755-2008, we consider the recent version of monthly sunspot 
number data as compiled by  "WDC-SILSO, Royal Observatory of Belgium, Brussels" (http://www.sidc.be/silso/). Before the year 1755, only yearly Belgium data 
is available. For this purpose, in order to keep the uniformity and as 
described in the following, this yearly data is not suitable for getting
the very good fit from equation (1). Hence, for back cast of activity 
of solar cycles, before the year 1755 AD, for the years 1700-1755, 
modeled total irradinace (kindly provided by Dr. Lean, Naval Observatory, USA;
Lean and Brueckner 1989; Lean 1990; Lean 1991) 
data is used for reconstruction of the sunspot data. 

Following the previous study (Hiremath 2006), monthly sunspot data
is normalized as follows. If $x_{i}$ are monthly means of sunspot number, ${\bar x}$ is the cycle
mean and $\sigma$ is the standard deviation, then the normalized deviation of sunspot data
is $y_{i}=(x_{i}-\bar x)/\sigma$. With the Levenberg-Marquardt algorithm (Press {\em et.al} 1992),
 such a normalized sunspot data is subjected to non-linear
least square fit for the following solution of a forced
and damped harmonic oscillator

\begin{equation}
y=A_{1}\cos(\omega t - \phi_{1}) + A_{2}\cos(\omega' t - \phi_{2}){\rm e}^{-\gamma t} ,
\end{equation}

\noindent where $y$ is displacement (in the present context we consider  
sunspot number), $A_{1}$, $A_{2}$ are amplitudes, $\omega (2\pi/T$, where 
$T$ is the period in years) is the sinusoidal frequency, $\omega '$ ($2\pi/T'$,
 where $T'$ is the period in years) is the damping frequency, $\phi _{1}$ and 
$\phi _{2}$ are the phases, $\gamma $ is the decay factor and $t$ is the time 
variable in months. On RHS of above equation, first term is called the ``steady" 
and second term the ``transient" parts of solution of a forced and damped
harmonic oscillator (Tipler and Mosca 2003). For different solar cycles from 1-23, 
observed sunspot data (blue curve)  with the over plotted non-linear least square 
fit (red continuous line) is illustrated in Figures 1-3 respectively. 

\section{Results}

\subsection{ Steady and Transient parts During Cycles 1-23}

For different cycles 1-23, Figure 4 illustrates the variation
of different physical parameters of the steady
and transient parts of solution of a forced and damped harmonic oscillator.
For the steady part, Fig 4(a) illustrates cycle to cycle variation
of amplitude $A_{1}$ (blue continuous line), frequency $\omega$ (red dotted line) and the
phase $\phi_{1}$ (red dashed line). Whereas physical parameters of transient part such as
amplitude $A_{2}$ (red continuous line) and decay factor $\gamma$ (dotted line) are
presented in Fig 4(c). Figure 4(d) illustrates frequency $\omega^{'}$ (red dotted  line) and 
phase $\phi_{2}$ (red dashed line).
In both the figures (4(c) and 4(d)), blue continuous line
represents the cycle mean (normalized to the maximum value in all the solar cycles).
 
As in the previous study (Hiremath 2006), one can notice
from three Figures (4(a), 4(c) and 4(d)) that frequency 
($\sim$ 22 yrs), amplitude and phase (except during few years) 
of the steady part
of the solar oscillator remains almost constant. 
Hence, an inevitable conclusion, as in the previous study (Hiremath 2006), is that
there is a constant perturber in the deep interior that creates
and maintains the solar cycle and activity phenomena with near
periodicity of 22 yrs. 

In the previous (Hiremath 2006) study, we find that whenever difference in phase
of steady and transient parts reaches a maximum, solar cycle and
activity phenomena on longer time scales ( $\sim$ 100 yrs) attains
a deep minimum, although not as deep as so called Maunder
minimum type of activity. For both the data set, in Fig 4(b), phase difference 
($\phi_{1}-\phi_{2}$) between the steady and transient parts is illustrated.
One can notice that nearly for every 100 years, sunspot activity
is  minimum.

\begin{figure}
{\bf \hskip 5ex Fig-5(a)} \hskip 50ex  {\bf Fig-5(b)}
\centerline{\psfig{figure=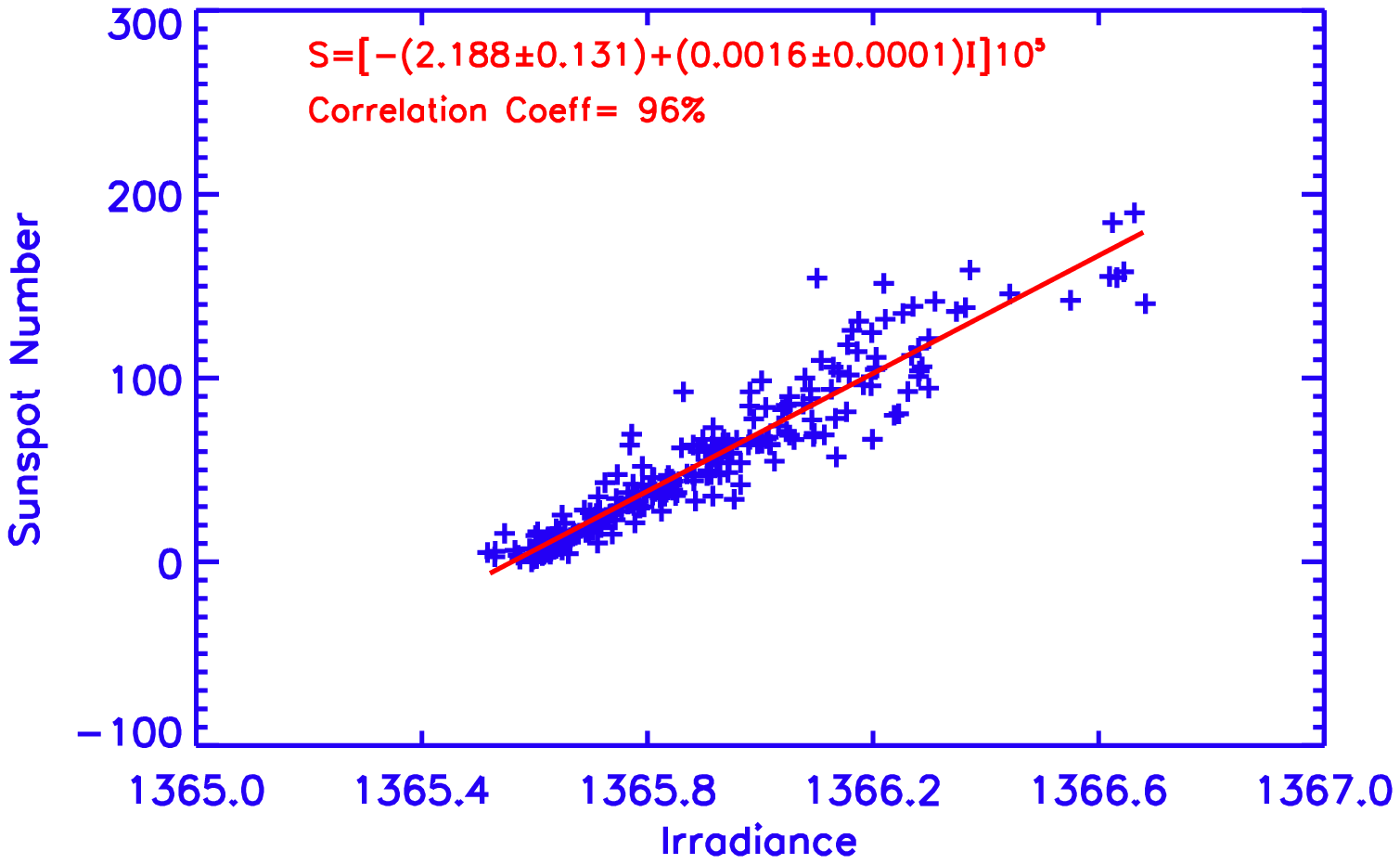,width=8cm,height=6cm}
\psfig{figure=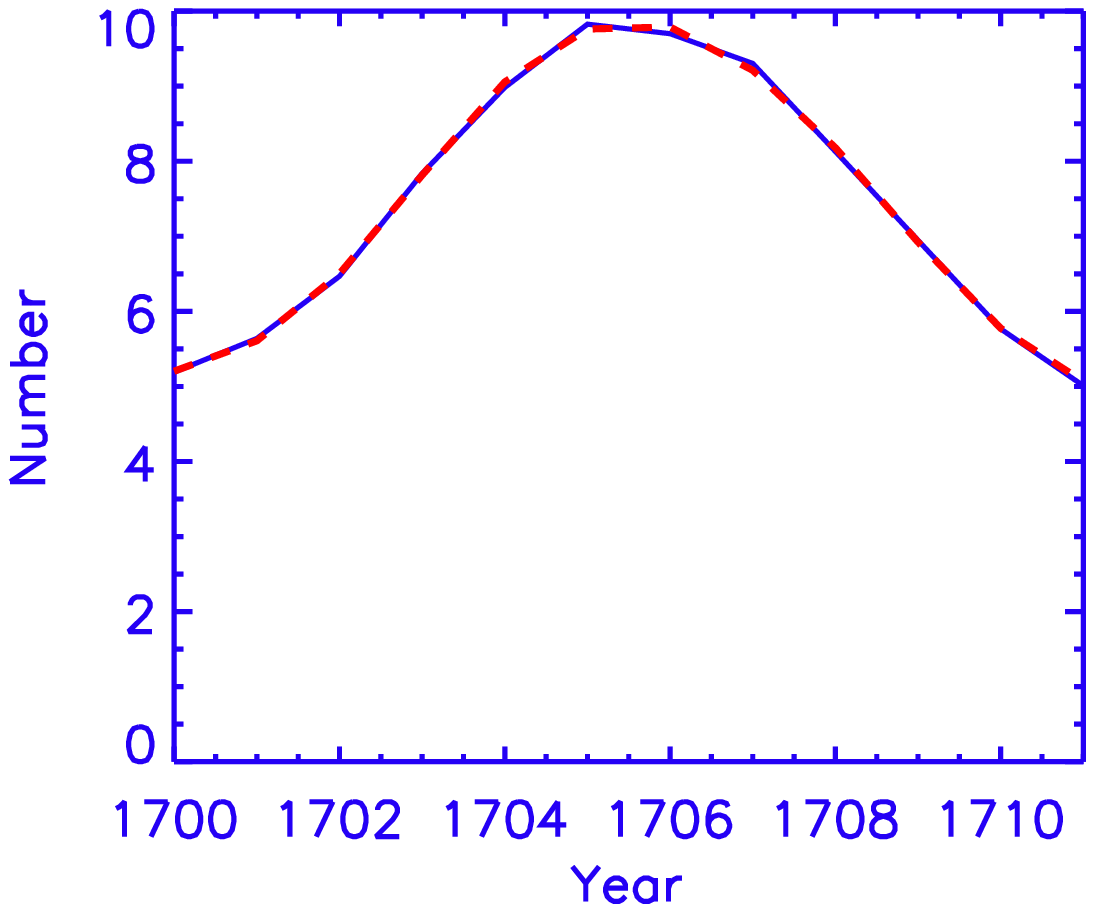, width=8cm,height=6cm}}
{\bf \hskip 5ex Fig-5(c)} \hskip 50ex  {\bf Fig-5(d)}
\centerline{\psfig{figure=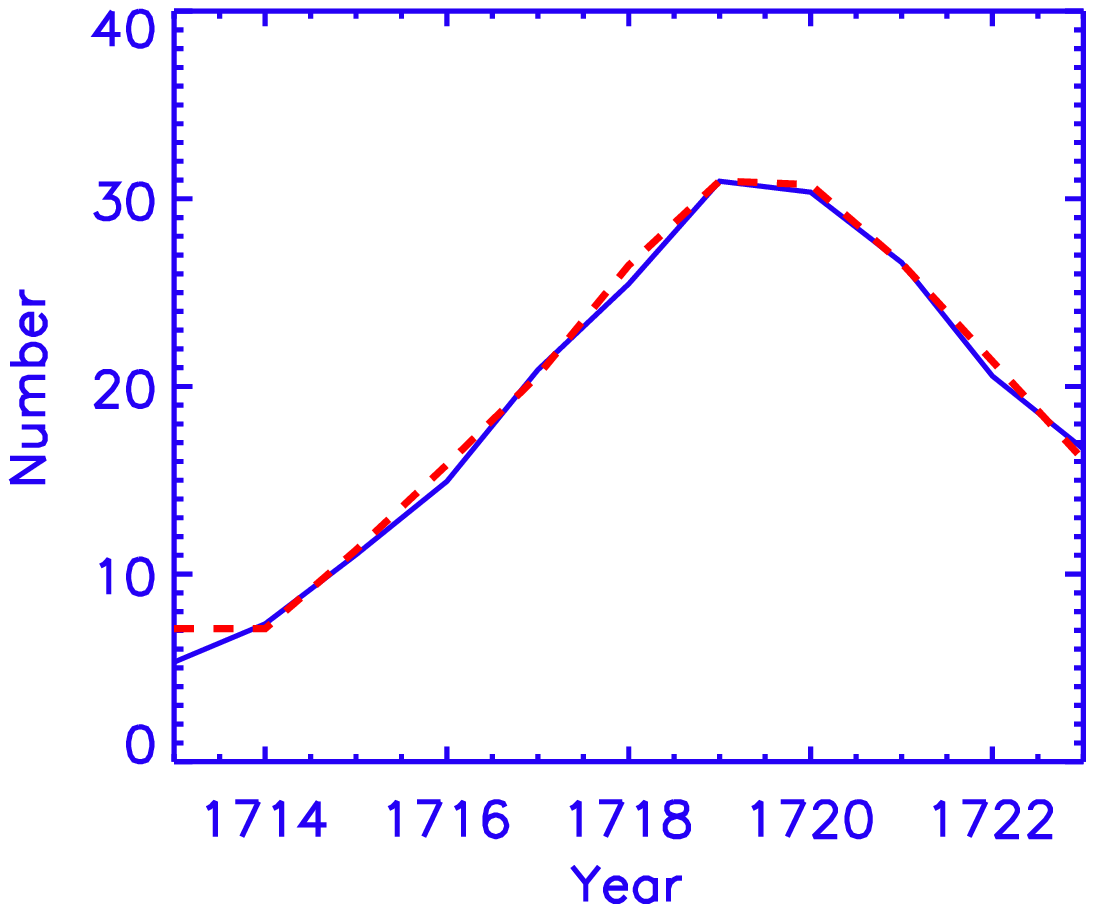, width=8cm,height=6cm}
\psfig{figure=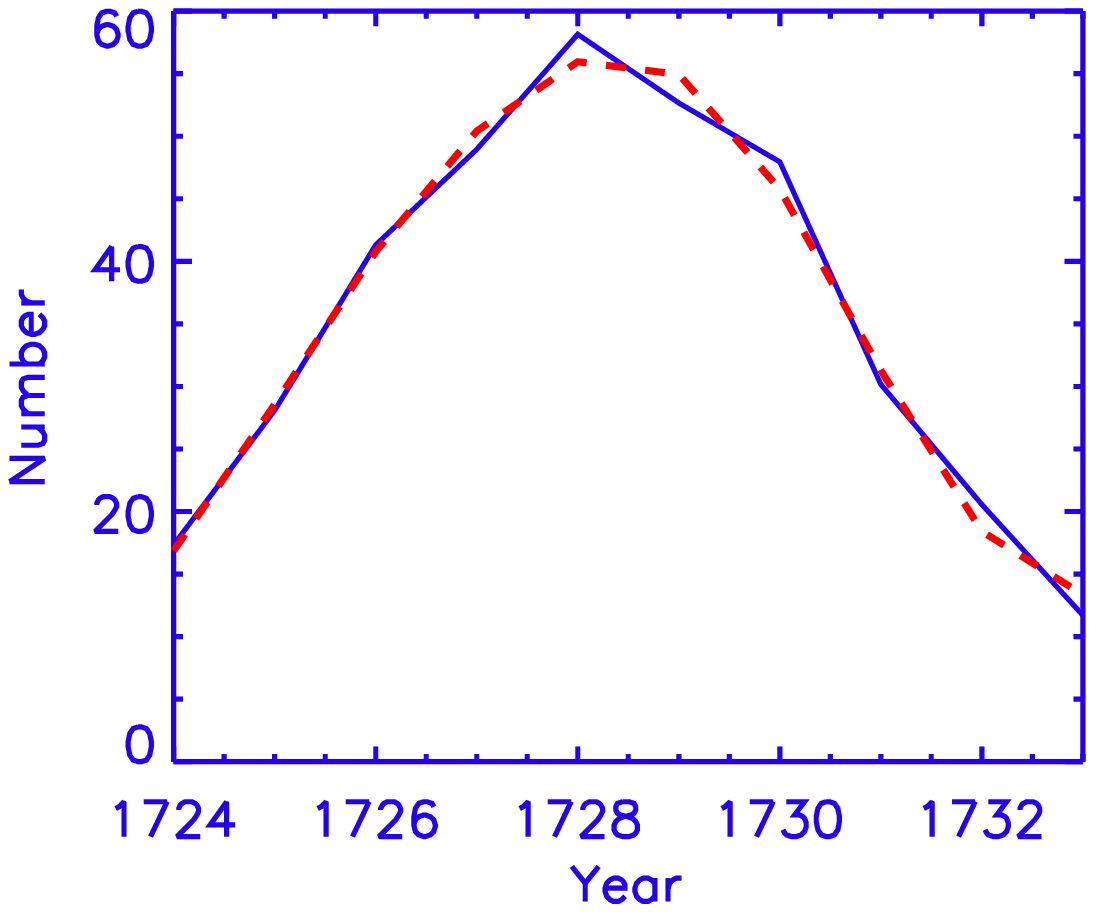, width=8cm,height=6cm}}
{\bf \hskip 5ex Fig-5(e)} \hskip 50ex  {\bf Fig-5(f)}
\centerline{\psfig{figure=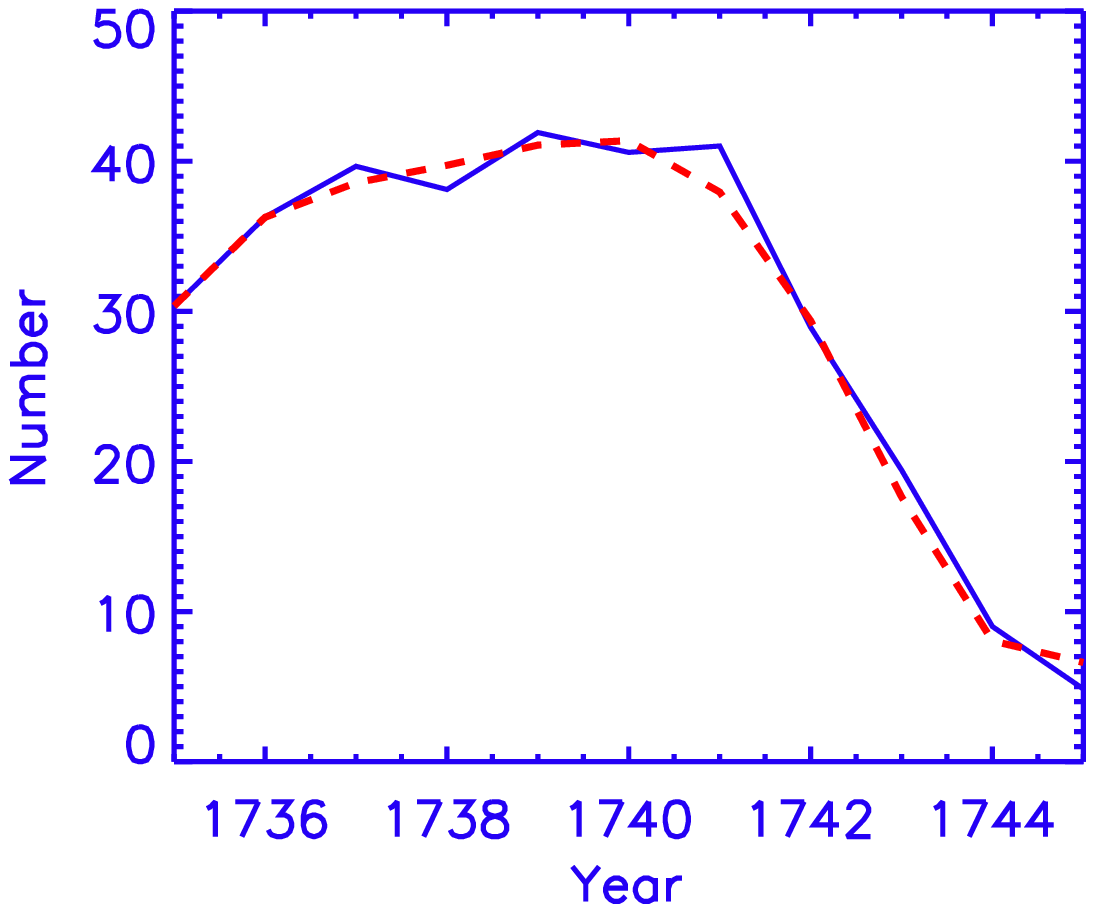, width=8cm,height=6cm}
\psfig{figure=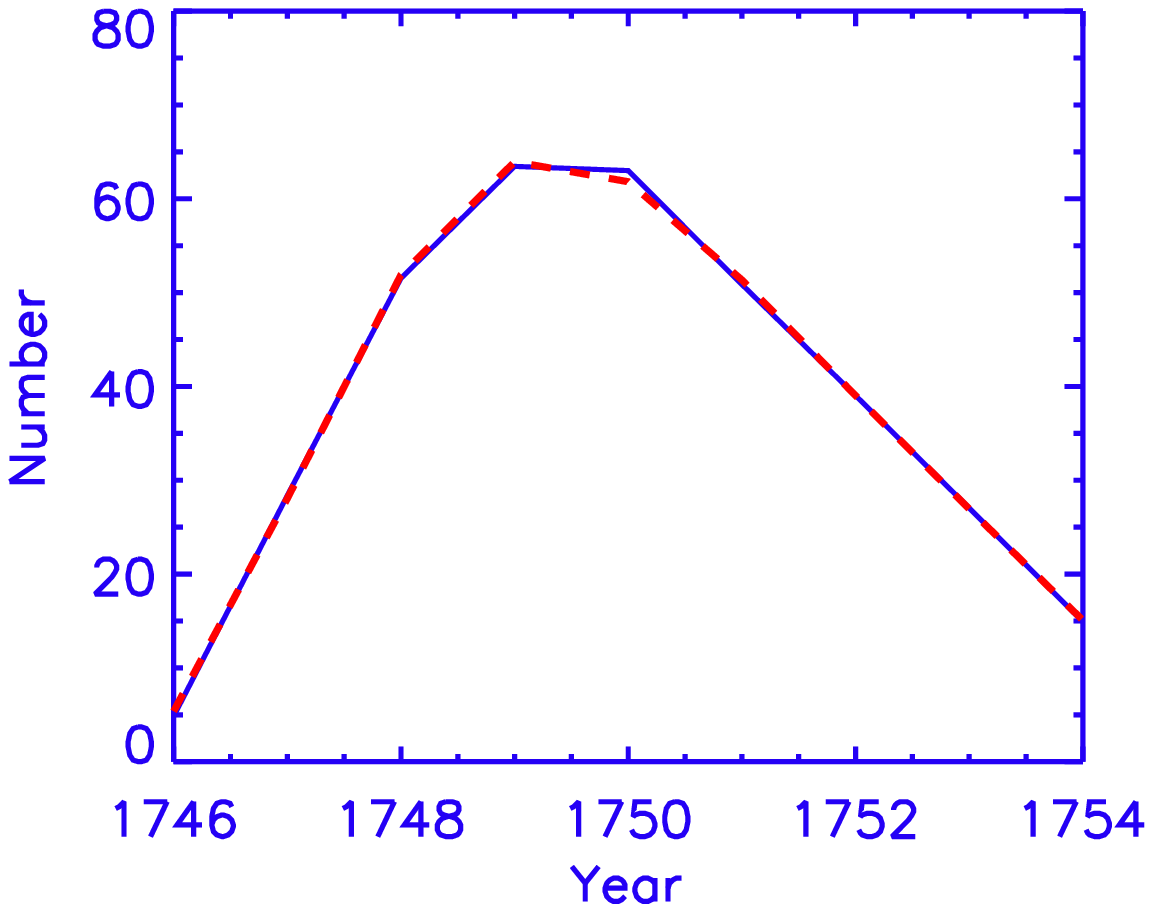, width=8cm,height=6cm}}
\caption{Fig 5(a) illustrates a scatter plot between the total irradiance
and the sunspot number. Overplotted red continuous line represents the 
least square fit between two variables. Whereas, for the cycles -5 to -1,
 Fig-5(b) to Fig-5(f) represent the monthly mean variation of Lean's total
irradiance (blue continuous line) data overplotted (dashed red line)
with the reconstructed sunspot data.}
\end{figure}

\begin{figure}
\centerline{\psfig{figure=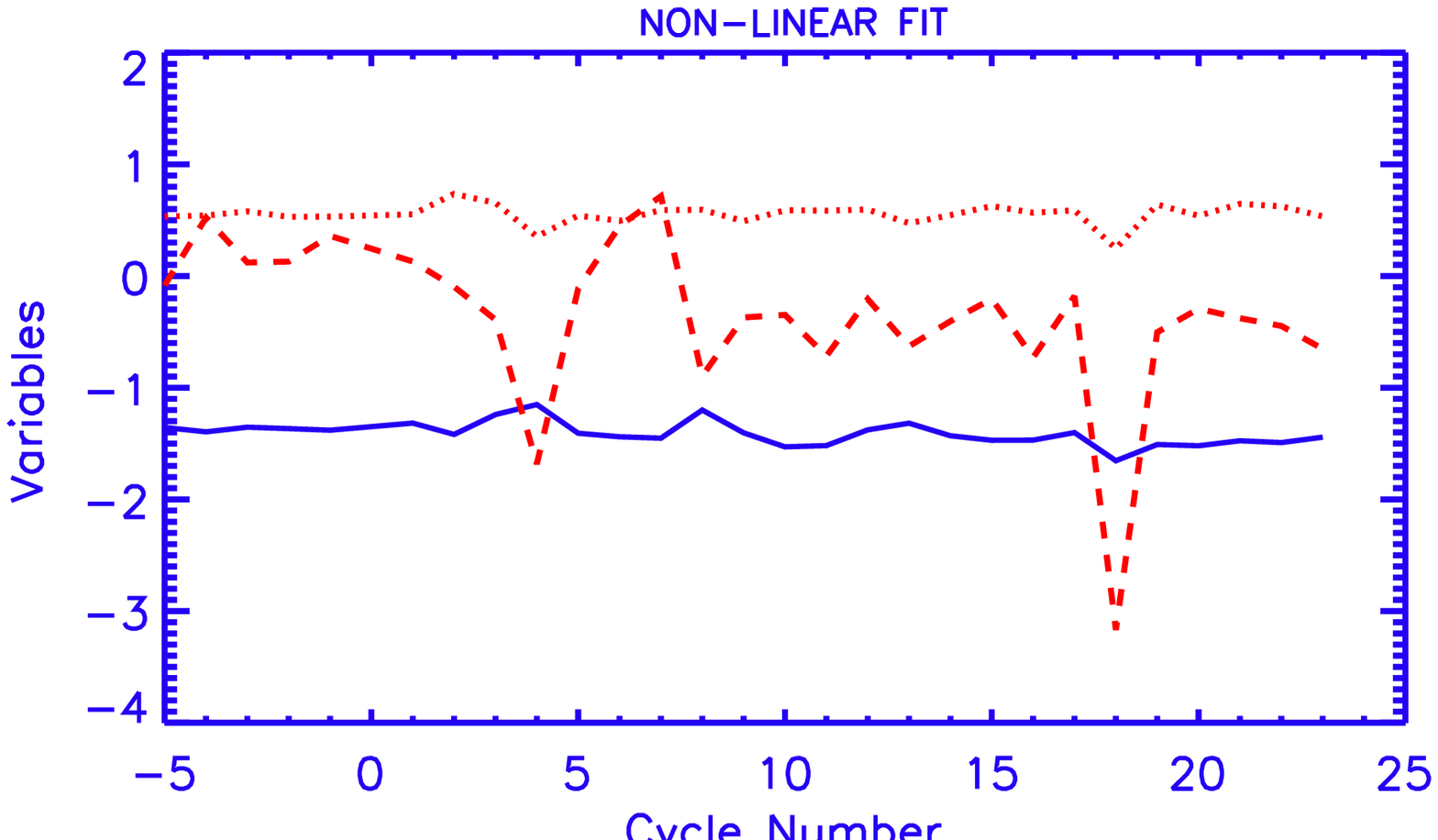,width=8cm,height=6cm}}
{\bf \hskip 5ex Fig-6(a)} \hskip 50ex  {\bf Fig-6(b)}
\centerline{\psfig{figure=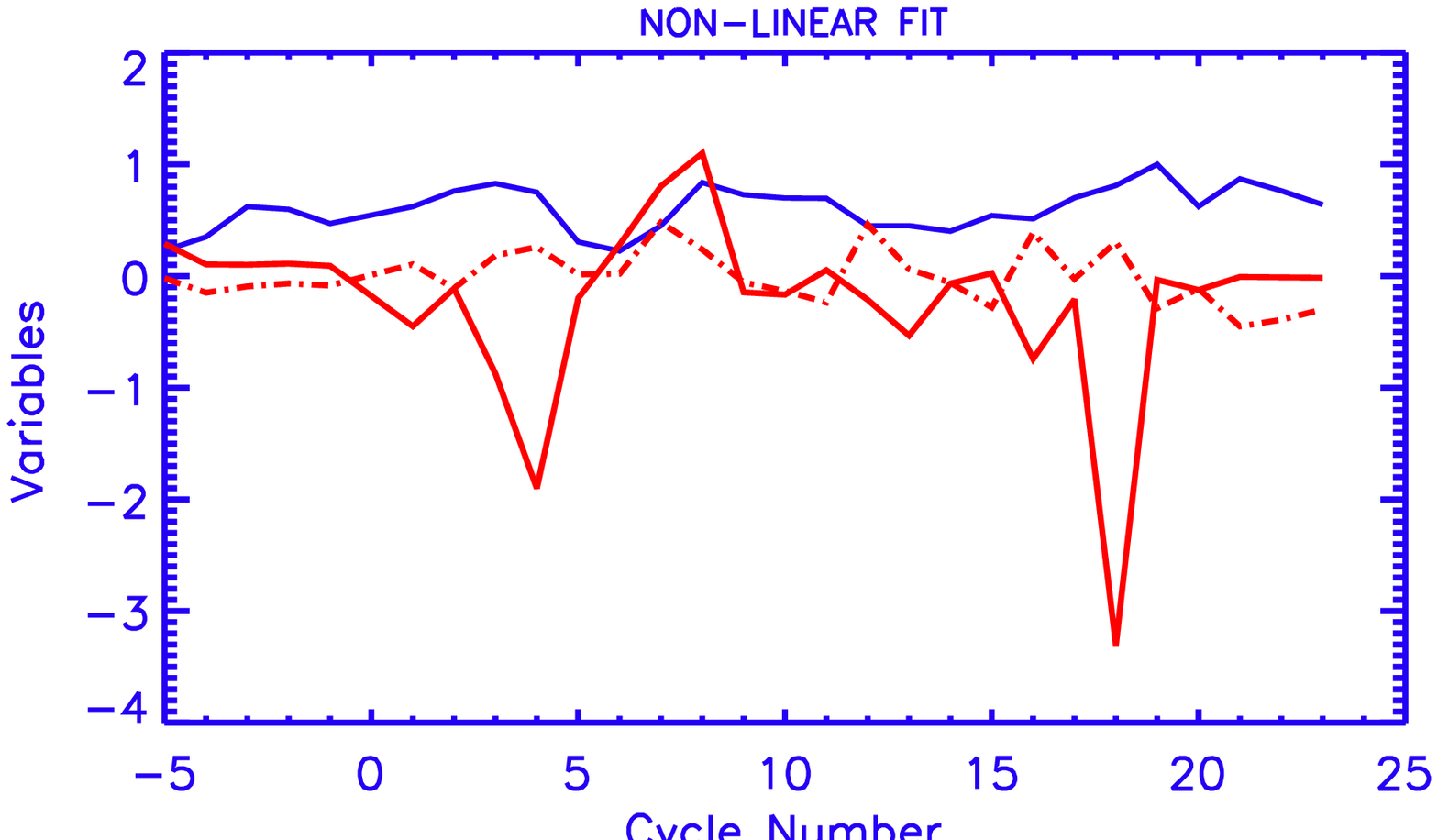, width=8cm,height=6cm}
\psfig{figure=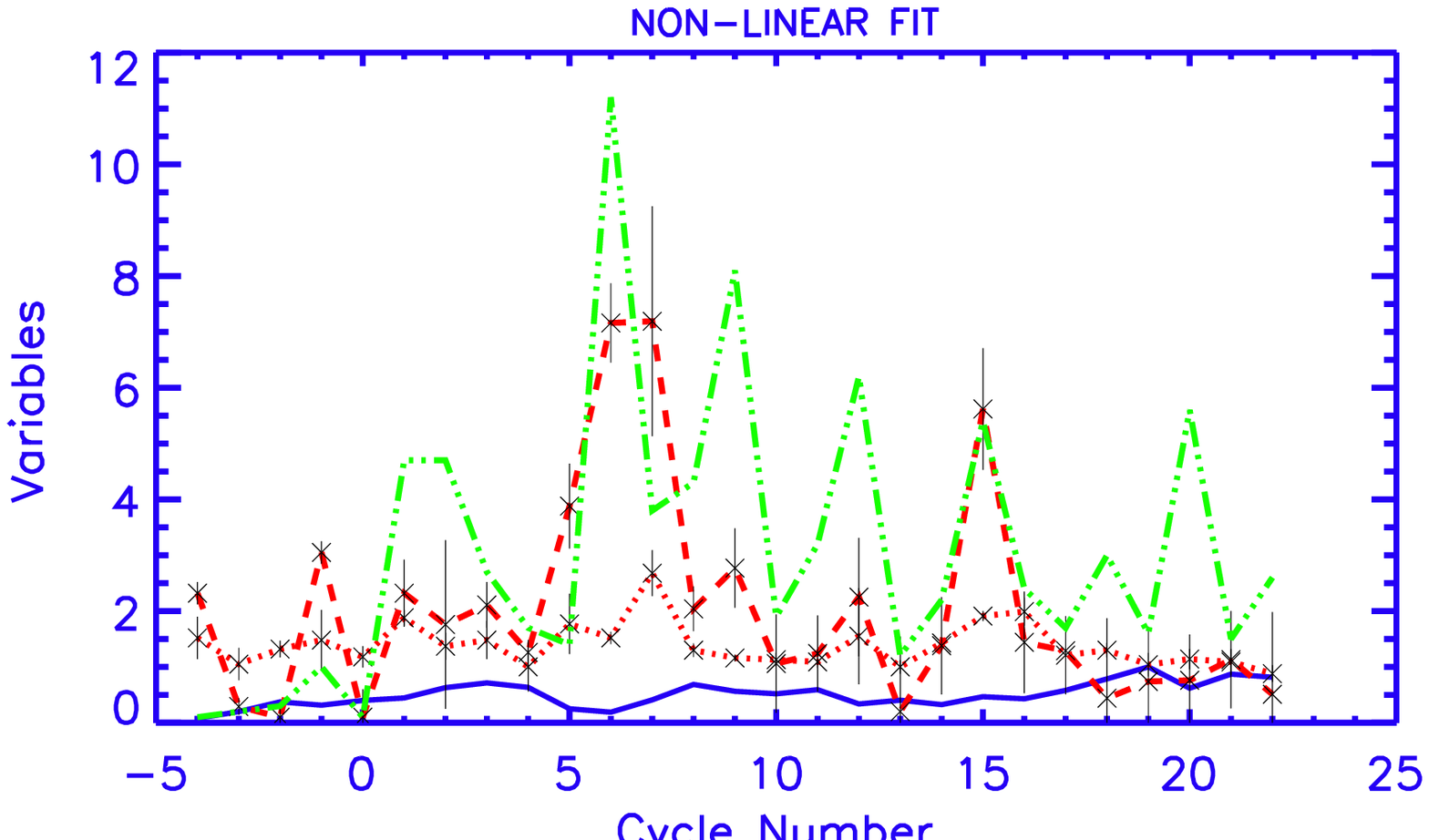, width=8cm,height=6cm}}

\caption{ For the extended cycles (-5 to 23), a nonlinear least square fit 
of a solution of forced and damped harmonic oscillator. 
For the extended solar cycles -5 to 23, upper panel of Fig 6 illustrates the variation
of different coefficients (blue continuous line is the amplitude $A_{1}$,
the dotted line is the frequency $\omega$ and the dashed line is the
phase $\phi _{1}$) for the steady part. Whereas variation of different physical
parameters of transient part is illustrated in Fig 6(a) and 
Fig 6(b) respectively.
In Fig 6(a), red continuous line is the amplitude $A_{2}$ and the dash with
dotted line represents the decay factor $\gamma $ respectively.
In Fig 6(b), the dotted line is the frequency $\omega '$ and the dashed
line is the phase $\phi _{2}$. The dash with three dotted line represents
the values of $\chi ^2$ (a measure of goodness of fit) for each cycle.
For all the cycles -5 to 23, in both the figures Fig-6(a) and Fig-6(b), 
blue continuous line is cycle mean of
sunspot number.}
\end{figure}

\subsection{Activity Before the Era of Cycles 1-23}

Let us examine whether all the physical parameters of
steady part of solution of a forced and damped harmonic oscillator 
are constant or not during the 16th century. Another investigation is to 
confirm whether sun really experienced
a Maunder type of deep minimum or not. As there are no systematic sunspot
observations before the era of cycle 1, we follow the following
method in order to reconstruct the sunspot data. For the years 1700-1755, Lean's
total irradiance data is considered, and different cycles minima
are estimated. Before the era of regular sunspot data that
starts with cycle 1, first previous cycle data is assigned as -1 cycle, 
next cycle -2 and so on. 

For the period of 1755-2008, sunspot data from the regular observations
is considered and association is examined with the
Lean's reconstructed TSI data for the same period. We find that both the variables are
strongly positively correlated and the estimated correlation is 96\%. Such an association and hence a scatter
plot between two variables is illustrated in Fig 5(a). Both the
variables are also subjected to a linear least square fit and the
following law between the two variables is obtained

\begin{equation}
S=[-(2.188 \pm 0.131) + (0.016 \pm 0.000096)I] 10^{5} ,
\end{equation}
\noindent where $S$ is the sunspot activity and $I$ is reconstructed total radiance.
With this empirical relation, sunspot data is reconstructed during 1700-1750. As described in the 
previous section, with a solution of forced and damped harmonic oscillator, 
such a reconstructed sunspot activity is subjected to a
nonlinear least square fit and different physical parameters are estimated
For the cycles -1, -2, -3, -4, -5,  Figures 5(b)-5(f)  illustrate the reconstructed
sunspot activity (blue continuous line) with over plotted fit (red dashed line). 
Whereas Fig 6 represents different physical
parameters of steady and transient parts of a forced and damped
harmonic oscillator. One can notice that, as in section 3.1, for all
the solar cycles (-5 to 23), frequency ($\sim$ 22 yrs) and amplitude of the steady part
remains approximately constant and amplitude of the transient part is 
phase locked with the steady part.

\begin{figure}
\centerline{\psfig{figure=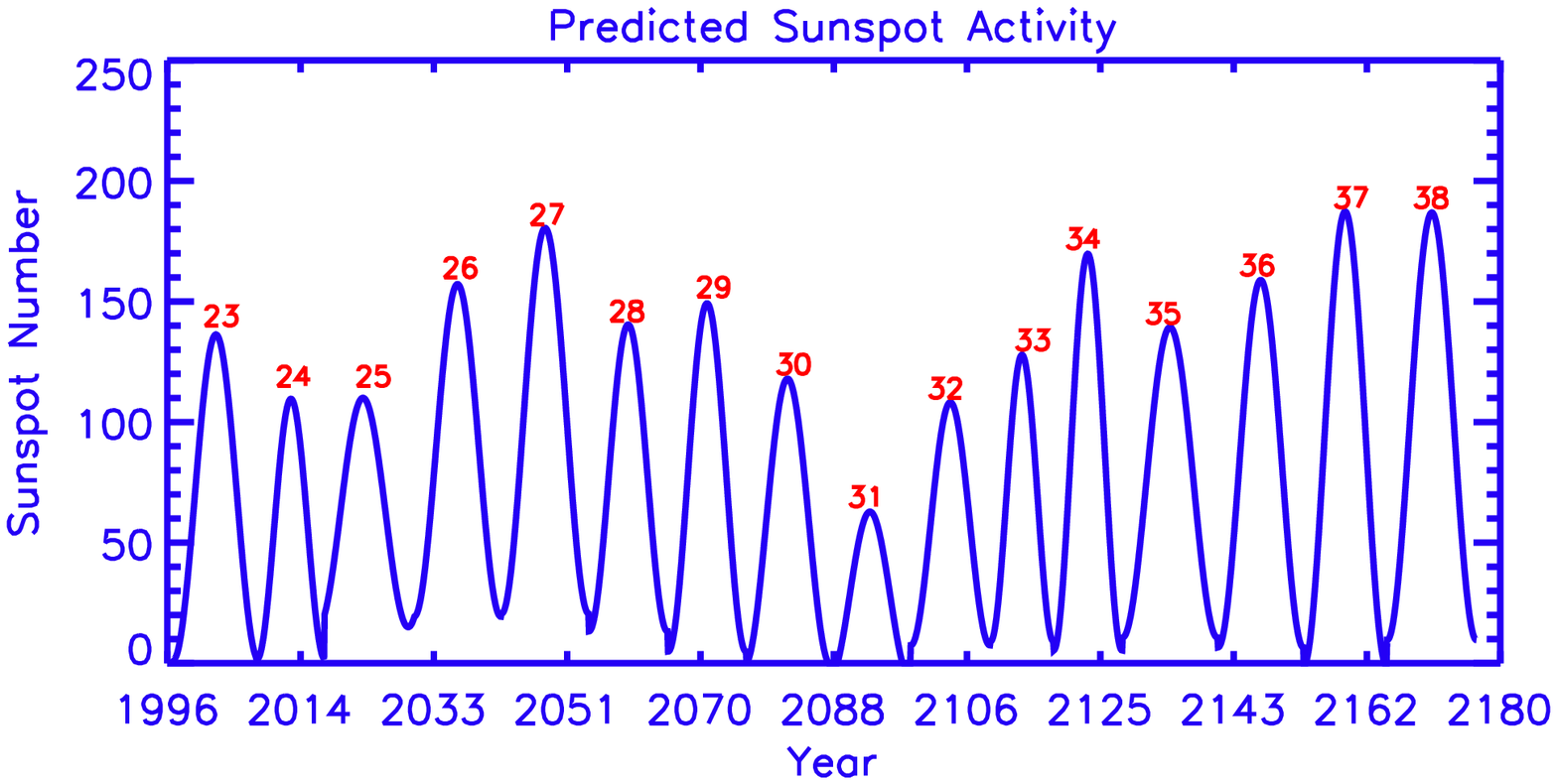,width=15cm,height=6cm}}
\centerline{\psfig{figure=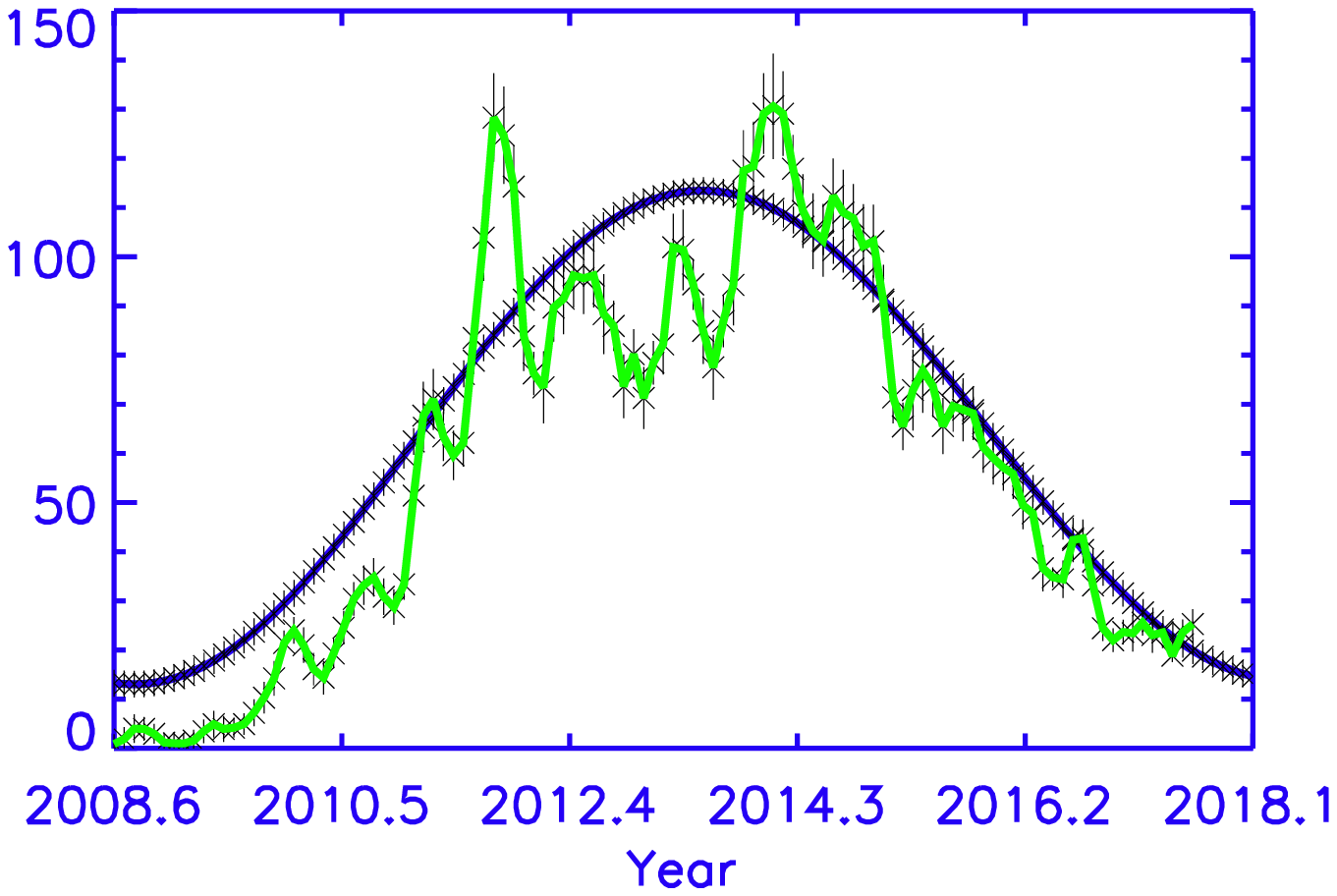,width=8cm,height=6cm}
\psfig{figure=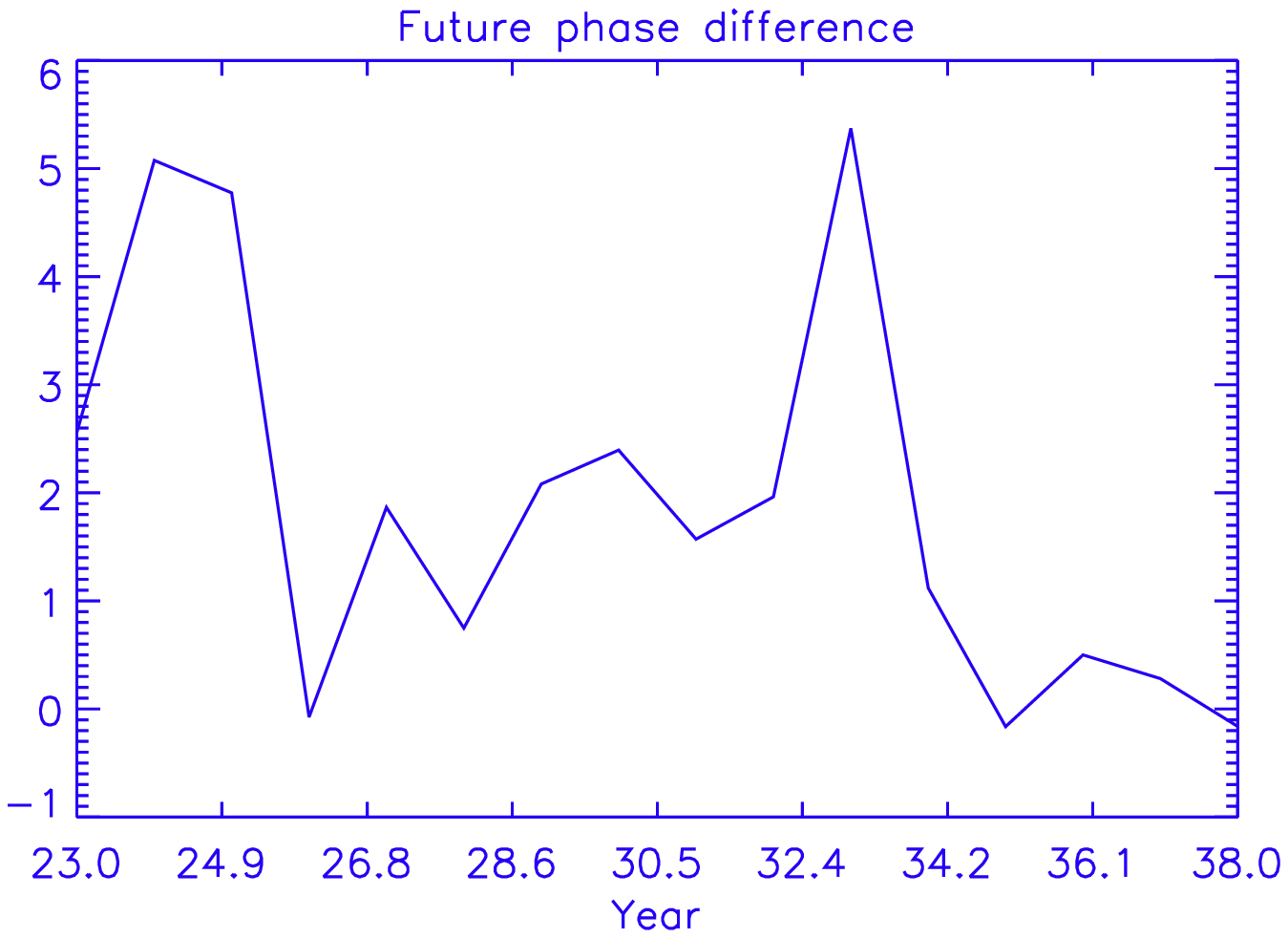, width=8cm,height=6cm}}
\caption{For the solar cycles 23 and beyond forecast of amplitudes of 
future 16 solar cycles (upper panel). For typical cycle 24, first figure
of lower panel illustrates the monthly 
variation of predicted (blue continuous line with error bars) and the 
observed (green continuous line with error bars) sunspot data.  Where as second 
figure in the lower panel  illustrates the predicted phase difference 
between the steady and transient parts for the future cycles.}
\end{figure}

\begin{figure}
{\bf \hskip 5ex Fig-8(a)} \hskip 40ex  {\bf Fig-8(b)}
\centerline{\psfig{figure=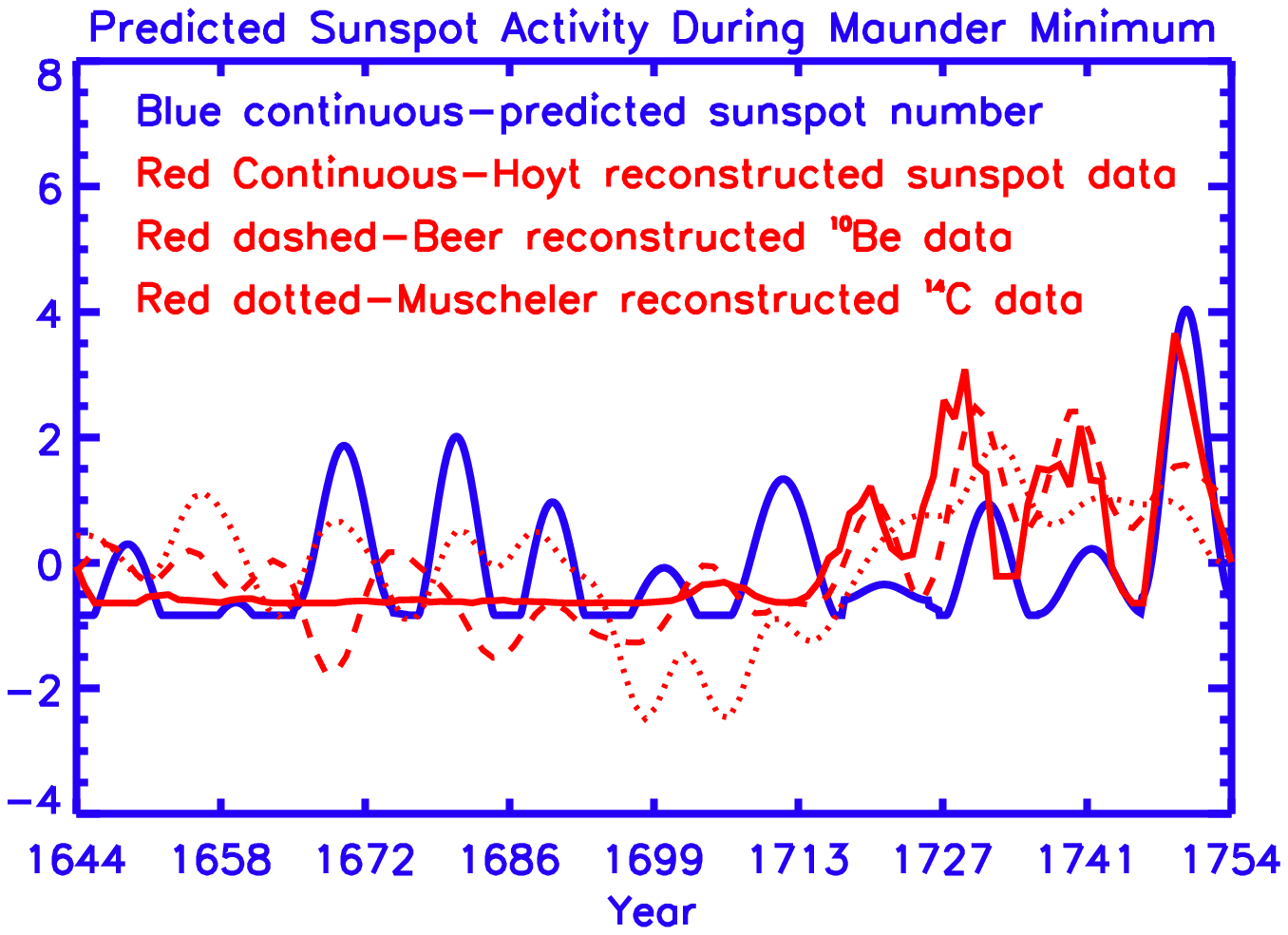,width=8cm,height=8cm}
\psfig{figure=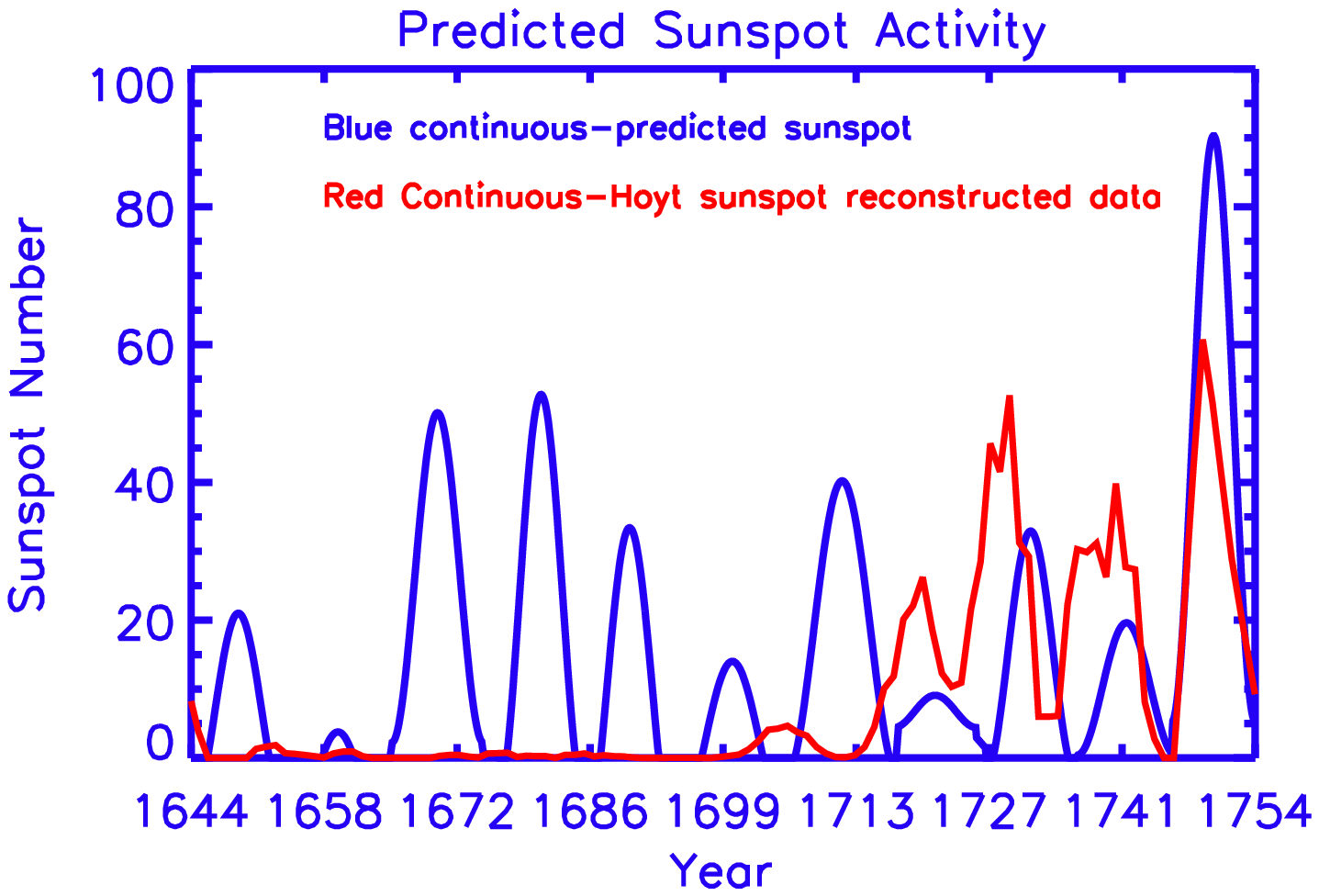, width=8cm,height=8cm}}
\caption{For the solar cycles -1 to -5 and beyond upto 1640 AD, backcast of the solar activity. Fig 8(a) illustrates
the deviation (from the mean and normalized with their respective
standard deviations) of different solar and plaeoclimatic activities. Blue continuous
line is sunspot activity obtained from the backcast. Red continuous line is
Hoyt's reconstructed sunspot data. Red dashed line is reconstructed $Be10$ data
from Beer {\em et.al.} (1998) and, red dotted line is reconstructed $C14$ data from 
Muscheler {\em et.al.} (2007).  
For the same cycles, whereas Fig 8(b) illustrates the absolute (without normalization) 
sunspot (blue continuous line) data from backcast and, Hoyt reconstructed
sunspot data (red continuous line).}
\end{figure}

\subsection {Forecast of the Solar Cycles}

There are many studies (Quassim, Attia and Elminir 2007; 
 Bhatt, Jain and Aggarwal 2009; 
Uwamacharo and Cilliers 2009; Volobuov 2009; 
Kilcik {\em et.al.} 2009;  Kane 2010, Baolin 2011; Kakad 2011; Nobel and Whatland 2012; 
Du 2012; Pesnell 2012; Du  2012; Yu et.al. 2012; Uzal, Piacentini and Verdes 2012;
Attia and Hassan 2013; Kane 2013; 
Helal and Gelal 2013; Pishkalo 2014; Kilcik {\em et.al.} 2014; 
  Li, Feng and Li 2015; Janardhan {\em et.al.} 2015; Raschna and Sarychov 2015;
Gao 2016; Hathaway and Upton 2016; Deminov, Nepomnyaschaya and Obridko 2016; 
Kitashivili 2016; Travaglini 2017; Gopalswamy {\em et.al.} 2018; Pesnell 2018; Pesnell and Schatten 2018; 
Sabarinath and Anilkumar 2018; Okoh {\em et.al.} 2018; Petrovay {\em et.al.} 2018; Singh and Bhargawa 2019;
Pala and Atici 2019; Okoh and Okoro 2020; Miao {\em et.al.} 2020; Wu and Qin 2021; Velasco {\em et.al.} 2021;
Maddanuand and Proietti 2022; Du 2022; Dang {\em et.al.} 2022) that forecast amplitudes of cycles 24 and 25. There are
also excellent reviews (Petrovay 2010; Pesnell 2012; Lopes {\em et.al.} 2014; Tripathi 2016; Petrovay 2020; Nandy 2021)  on the prediction of 
solar cycle. Whereas forecast of
 solar cycles beyond 24 and 25 are rare. In addition, forecast of cycle periods is
at most importance for the future space missions and space weather predictions.


Using sunspot data and autoregressive model, for the years of 1755-1996, in the previous study (Hiremath 2006),
physical parameters of steady part of forced and damped harmonic  
were used to forecast the amplitudes and periods of steady part. 
In this study, as we have extended sunspot data set from
 1700-2008, whole of this data set is used for estimation of different physical parameters
of steady and transient parts. With these extended physical
parameters and the autoregression model (see the description as given in section 1, 
page 46, second column, Hiremath 2008),
physical parameters of steady and transient parts are estimated for the
 16 cycles (cycles 23-38). With the predicted physical
parameters, monthly sunspot data is reconstructed from the physical parameters of steady
part as this term mainly reproduces the amplitude and period of a solar cycle, 
Top panel of Fig 7 illustrates the temporal variation of amplitudes of solar cycles 23 and beyond.
For the sake of comparison with the observed sunspot data,
for typical solar cycle 24, first figure of lower panel of Fig 7
 illustrates the
predicted (blue continuous line with error bars) amplitude
of the solar cycle. We find the stunning
result that not only the amplitude but also the period of the
predicted cycle 24 nearly matches very well with the observed amplitudes and period of
the solar cycle.

Another interesting part of our study is that
whenever the difference between the phases of steady and transient parts attain
the maximum, for every 100 years, few cycles around that period 
have minimum sunspot activity, 
although it is not as deep minimum as so called Maunder minimum type activity.
For example (see upper and lower panels of Fig 7) around 2100 AD, 
amplitude of sun's activity probably reaches
very low and minimum. However, notice that the predicted sunspot activity
 beyond cycle 24, as argued by some studies (Janardhan {\em et.al.} 2015;
 Sancez-Sesma 2016; Zharkova {\em et.al.} 2015), will never attains a 
Maunder minimum type of solar activity at least upto 200 yrs.

\subsection{Backcast of the Solar Cycles}

Using same physical parameters of steady part of forced and damped harmonic
oscillator for the years 1700-2008, with the autoregressive model, we backcast
the amplitude of the solar cycle upto the years 1645. Idea of this exercise
is to examine whether sun really experienced a deep Maunder minimum (during
the period of 1645-1700 AD) type of solar activity.
Figure 8 illustrates the backcasted solar activity. 
Fig 8(a) illustrates
the deviation (from the mean and normalized with their respective
standard deviations) of different solar and plaeoclimatic activities. Blue continuous
line is sunspot activity obtained from the backcast. Red continuous line is
Hoyt's reconstructed sunspot data. Red dashed line is reconstructed $Be10$ data
from Beer {\em et.al.} (1998) and, red dotted line is reconstructed $C14$ data from
Muscheler {\em et.al.} (2007).
For the same cycles, whereas Fig 8(b) illustrates the absolute (without normalization)
sunspot (blue continuous line) data from backcast and, Hoyt reconstructed
sunspot data (red continuous line).
Of course, one can notice from these figures that, although we find 
a dearth of solar activity around 1650-1662, our backcast 
sunspot data did not reproduce
so called  deep Maunder minimum of solar activity during 1600-1645. Infact, these
results are also consistent with the previous (Komitov and Kaftan 2003; 
Hiremath 2010; De Jager and Duhau 2012; 
Steinhilber and Beer 2013; Velasco Herrara {\em et. al.} 2015; 
Gao 2016; Svalgaard and Schatten 2016; Travglini 2017) studies
which do not agree with the idea that there was a dearth of sunspot
activity in particular and sun's other activity (such as coronal
holes, coronal mass ejections, flares etc.,) in general
(see Feyman 1982; Legrand and Simon 1991; Cliver, Boriakoff and
Bounar 1998; Feyman 1982; De Jager 2005; Georgieva and Kirov
2006)  around 1645-1700 AD as claimed by Maunder. 
However, further analysis is required to confirm this result 
as linear autoregressive method has certain limitations.

\section{Conclusions}
Shape of the solar cycle is described as a solution of a forced and damped
harmonic oscillator. First, for the years 1700-2008 (cycles
1-23), updated monthly sunspot data is subjected to a nonlinear 
least square fit of this solution and
different physical parameters (amplitudes, phases and frequencies) of steady
and transient parts of the solar oscillator are estimated. Following are
the important findings: (i) for all the cycles -5 to -1 and 1-23, 
amplitude, frequency ( or period of $\sim$ 22 yrs) 
and the dissipation factor $\gamma$ remain approximately constant,
(ii) amplitude of the transient part is phase locked with the phase
of the steady part, (iii) whenever phase difference between
the steady and transient parts attains a maximum, sunspot
activity also reaches a minimum, although this minimum is
not as deep as so called Maunder minimum and, (iv) for the 
period 1755-2008, we find a linear relationship (with
a high correlation) between the sunspot and total irradiance 
activities.  
 
Before the era of regular observations of sunspots, from 1700-1755, 
Lean's total irradiance data is considered and, by detecting different
minima, 5 cycles (-1 to -5) are obtained. From the obtained
relationship between sunspot and total irradiance data, sunspot
number data is reconstructed for all the 5 cycles (1700-1755). 
All these five cycles sunspot  
data is subjected to a nonlinear least square fit of a
forced and damped harmonic oscillator and, different
physical parameters of steady and transient parts are
estimated. We find that cycle to cycle variation of all the physical
parameters estimated from these 5 cycles data is almost
same as cycle to cycle variation of physical parameters
estimated from 1-23 cycles.

These 5 cycles physical parameters are
merged with the physical parameters of cycles 1-23
estimated from the systematic sunspot observations. 
Finally, with an autoregressive model, 
different physical parameters that are estimated
from the combined data set are used for forecast
and back cast of the amplitude and periods of the
solar cycles. Important results are: (i) predicted
amplitude and period of cycle 24 are almost
same as the amplitude and period of the observed sunspot cycle,
(ii) amplitude of upcoming solar cycle 25 is almost
same as present 24th cycle, (iii) as expected 
by other studies, there is
no imminent so called deep Maunder minimum type
of solar activity in future, at least upto 200 yrs
and, (iv) backcasted sunspot activity suggests that 
although there are intermittent dearth
of sunspot activity, we did not find a very long (1645-1700)
deep minimum of solar activity as claimed by Maunder. 
\vskip 0.2cm


\vfill\eject

\end{document}